\documentclass[proof]{pasj00}
\usepackage{multirow}

\begin{document}
\SetRunningHead{Murakami et al.}{Calibration of the AKARI/FTS}
\Received{2010/03/14}%{yyyy/mm/dd}
\Accepted{2010/06/14}%{yyyy/mm/dd}

\title{Calibration of the AKARI Far-Infrared Imaging 
	Fourier Transform Spectrometer}

%%% begin:list of authors
%%%  Do NOT capitalize all letters in "textsc".
 \author{
   Noriko \textsc{Murakami}\altaffilmark{1,2},
   Mitsunobu \textsc{Kawada}\altaffilmark{2,4},
   Hidenori \textsc{Takahashi}\altaffilmark{3},
   Yoko \textsc{Okada}\altaffilmark{4,5},
   Akiko \textsc{Yasuda}\altaffilmark{2,4},
   Takafumi \textsc{Ootsubo}\altaffilmark{4,6},
   Hidehiro \textsc{Kaneda}\altaffilmark{2},
   Hiroshi \textsc{Matsuo}\altaffilmark{7},
   Jean-Paul \textsc{Baluteau}\altaffilmark{8},
   Peter \textsc{Davis-Imhof}\altaffilmark{9},
   Brad G. \textsc{Gom}\altaffilmark{10},
   David A. \textsc{Naylor}\altaffilmark{11},
   Annie \textsc{Zavagno}\altaffilmark{8},
   Issei \textsc{Yamamura}\altaffilmark{4},
   Shuji \textsc{Matsuura}\altaffilmark{4},
   Mai \textsc{Shirahata}\altaffilmark{4},
   Yasuo \textsc{Doi}\altaffilmark{12},
   Takao \textsc{Nakagawa}\altaffilmark{4},
   and
   Hiroshi \textsc{Shibai}\altaffilmark{13}}

\altaffiltext{1}{Bisei Astronomical Observatory, 1723-70 Okura, Bisei-cho, Ibara-shi, Okayama 714-1411, Japan}
 \email{noriko@bao.go.jp}
\altaffiltext{2}{Graduate School of Sciences, Nagoya University,
	Furo-cho, Chikusa-ku, Nagoya 464-8602, Japan}
 \email{kawada@u.phys.nagoya-u.ac.jp}
\altaffiltext{3}{Gunma Astronomical Observatory, 6860-86 Nakayama, 
	Takayama, Agatsuma, Gunma 377-0702, Japan}
\altaffiltext{4}{Institute of Space and Astronautical Science, JAXA,
	3-1-1 Yoshinodai, Chuo-ku, Sagamihara 252-5210, Japan}
\altaffiltext{5}{I. Physikalisches Institut, Universit\"{a}t zu K\"{o}ln, 
        Z\"{u}lpicher Str. 77, 50937 K\"{o}ln, Germany}
\altaffiltext{6}{Astronomical Institute, Graduate School of Science,
        Tohoku university, Aramaki, Aoba-ku, Sendai, 980-8578, Japan}
\altaffiltext{7}{Advanced Technology Center, 
	National Astronomical Observatory of Japan,
	2-21-1 Osawa, Mitaka, Tokyo 181-8588, Japan}
\altaffiltext{8}{Laboratoire d'Astrophysique de Marseille, 
        UMR 6110 CNRS \& Universite de Provence, 
        13388 Marseille Cedex 13, France}
\altaffiltext{9}{Blue Sky Spectroscopy Inc., 
 	9-740 4th Avenue South, Lethbridge, T1J 0N9, Alberta, Canada}
\altaffiltext{10}{University of Lethbridge,
	Lethbridge T1K 3M4, Alberta, Canada}
\altaffiltext{11}{Institute for Space Imaging Science, 
        University of Lethbridge, Lethbridge T1K 3M4, Alberta, Canada}
\altaffiltext{12}{Department of General System Studies, 
	Graduate School of Arts and Sciences, 
	The University of Tokyo, 3-8-1 Komaba, Meguro-ku, 
	Tokyo 153-8902, Japan}
\altaffiltext{13}{Graduate School of Sciences, Osaka University,
	1-1 Machikaneyama, Toyonaka, Osaka 560-0043,Japan}

%%% end:list of authors

%% `\KeyWords{}' always has to be placed before `\maketitle'.
 %Do NOT move this preamble from here!

\maketitle

\begin{abstract}

The Far-Infrared Surveyor (FIS) onboard the AKARI satellite has a
spectroscopic capability provided by a Fourier transform spectrometer
(FIS-FTS). FIS-FTS is the first space-borne imaging FTS 
dedicated to far-infrared astronomical observations. 
We describe the calibration process
of the FIS-FTS and discuss its accuracy and reliability. The
calibration is based on the observational data of bright astronomical
sources as well as two instrumental sources. 
We have compared the FIS-FTS spectra with the spectra
obtained from the Long Wavelength Spectrometer (LWS) of the
Infrared Space Observatory (ISO) having a similar spectral coverage.
The present calibration method accurately reproduces the spectra of 
several solar system objects having a reliable spectral model.
 Under this condition the relative uncertainty of the  calibration of the 
continuum 
is estimated to be $\pm$15\% for SW, $\pm$10\% for 70--85 cm$^{-1}$ of LW, 
and $\pm$20\% for 60--70 cm$^{-1}$ of LW; 
and the absolute uncertainty is estimated to be $+35/-55\%$ for SW, 
$+35/-55\%$ for 70--85 cm$^{-1}$ of LW, 
and $+40/-60\%$ for 60--70 cm$^{-1}$ of LW.
These values are confirmed by comparison with theoretical models and 
previous observations by the ISO/LWS.

\end{abstract}

\section{Introduction}

Since the Earth's atmosphere is opaque at far-infrared wavelengths, 
we have to use air- or space-borne instruments for observing 
the universe at these wavelengths. 
In the last few decades, developments
in infrared and space technologies have provided great opportunities to 
astronomers. In 1983, the
first infrared astronomical satellite, IRAS \citep{Neugebauer84},
provided the first view of the infrared universe. Following the
success of this pioneering mission, the first Japanese infrared
astronomical mission, InfraRed Telescope in Space (IRTS;
\cite{Murakami96}) and the European Infrared Space Observatory (ISO;
\cite{Kessler96}) were operated in the late 1990s. 
These missions extended the photometric observations by IRAS to 
spectroscopic measurements, which provided a new and powerful tool for
probing astronomical properties. 

The far-infrared spectral region is dominated by thermal emission of the
interstellar dust and some prominent atomic fine structure
lines, such as [C\,{\footnotesize II}] (63.4 cm$^{-1}$, 158 $\mu$m), 
[N\,{\footnotesize II}] (82.1 cm$^{-1}$, 122 $\mu$m) and 
[O\,{\footnotesize III}] (113.2 cm$^{-1}$, 88 $\mu$m).  
The spectral lines indicate
the physical properties of the interstellar medium. For example,
the [C\,{\footnotesize II}] line is a major cooling line in the
photo-dissociation regions (PDRs; \cite{Tielens85}) and was detected
around the Galactic Plane and nearby galaxies. 
The Far-Infrared Absolute
Spectrophotometer (FIRAS; \cite{Mather93b}), mounted on the Cosmic Background 
Explorer (COBE; \cite{Mather93a}, launched in 1983) 
detected the line/continuum emission from the
interstellar gas/dust of the Galaxy as well as the cosmic microwave
background radiation (\cite{Mather90}).
The Far-Infrared Line Mapper (FILM) onboard IRTS surveyed 
the [C\,{\footnotesize II}] line with high sensitivity; however, the 
surveyed area was limited to 7\% of the entire sky \citep{Shibai96}. 
Two balloon-borne telescopes, 
the Balloon-borne Infrared Telescope (BIRT) and 
the Balloon-borne Infrared Carbon Explorer (BICE), 
efficiently mapped the Galactic 
Plane with the same line (\cite{Shibai91}, \cite{Nakagawa95}). 
A rocket-borne instrument detected the same line in the Lockman Hole region 
with very high sensitivity (\cite{Bock94}, \cite{Matsuhara97}).
Unfortunately, after
the expiration of the Long Wavelength Spectrometer of ISO (LWS;
\cite{Clegg96}) there were no space-based facilities for 
far-infrared spectroscopy. 
The Spitzer Space Telescope (SST; \cite{Werner04}) has
been extremely productive for infrared astronomy; 
however it lacks a far-infrared spectroscopic capability.

The Japanese infrared astronomical satellite AKARI
\citep{Murakami07} was launched in February 22, 2006 (JST). 
One of the two focal-plane instruments is the Far-Infrared Surveyor 
(FIS; \cite{Kawada07}).
The primary purpose of the FIS was to accomplish a fine,
unbiased photometric survey of the entire sky in four photometric bands 
covering the wavenumber range from 
55 cm$^{-1}$ to 200 cm$^{-1}$. 
In addition, the FIS also has a slow-scan observation mode (\cite{Shirahata09}) and 
a spectroscopic observation mode provided by a Fourier transform
spectrometer (FIS-FTS). FIS-FTS is a Martin-Puplett type interferometer 
(\cite{Martin69}) and is similar in optical design to 
the composite infrared spectrometer (CIRS; \cite{Kunde96}) 
on the Cassini satellite.
A unique feature of the FIS-FTS is its detector system. Two sets of 
two-dimensional photoconductive detector arrays are installed at 
both the output
ports of the interferometer, providing an imaging spectroscopic capability.
Therefore, the FIS-FTS is the first imaging FTS for far-infrared
astronomy in space. 
The principal advantage of the FIS-FTS over the ISO/LWS is the high 
observational efficiency due to its imaging capability. 
The SPIRE instrument (\cite{Griffin08}) of the 
Herschel Space Observatory (\cite{Pilbratt08}), launched 
in May 2009, also contains an imaging FTS 
and provides fine spectral images of the submillimeter sky
%\footnote{Herschel Science demonstration workshop - http://herschel.esac.esa.int/SDPDP}.
\footnote{Herschel Science demonstration workshop - http://herschel.esac.esa.int/SDP\_DP\_wkshop.shtml }.

\begin{table}
\caption{The specifications of FIS-FTS.}
\label{tbl:ftsspec}
\begin{center}
\begin{tabular}{ccc}
\hline
{\bf detector unit} & SW & LW\\
\hline
Spectral Coverage\footnotemark[$*$] & 85 -- 130 cm$^{-1}$ & 60 -- 88 cm$^{-1}$\\
Array format & $3\times 20$ & $3\times 15$\\
Pixel scale & \timeform{26.8''} &  \timeform{44.2''}\\
FWHM(major/minor) & \timeform{44''}/ \timeform{39''} & \timeform{57''}/ \timeform{53''}\\
Sampling rate & 170.66\,Hz & 85.33\,Hz\\
\hline
{\bf operation mode} & SED mode & full-res. mode\\
\hline
OPD range (L)& $\pm 0.42$\,cm & $-0.9 \sim +2.7$\,cm\\
Resolution (1/2L$_{max}$) & 1.2\,cm$^{-1}$ & 0.18\,cm$^{-1}$\\
Mirror scan speed & $\sim$ 0.073\,cm s$^{-1}$ & $\sim$ 0.076\,cm s$^{-1}$\\
\hline
\multicolumn{3}{@{}l@{}}{\hbox to 0pt{\parbox{120mm}{\footnotesize 
	   \footnotemark[$*$] effective spectral coverage in this calibration.
	 }\hss}}
\end{tabular}
\end{center}
\end{table}

The FIS-FTS performance in space was generally consistent 
with that expected before launch laboratory tests. The
instrument had worked well until the loss of its liquid helium
on August 26, 2007 rendered the detectors inoperative.
During the on-orbit operation of one and a half years, the FIS-FTS achieved
600 pointed observations. One of the challenges in
analyzing the FIS-FTS data is the unique issues that arise from 
using photoconductive detector arrays in conjunction with an FTS.
In this paper, we describe how the incident spectra are reproduced
from the measured signals and discuss the accuracy and
reliability of the calibration. 
The next section introduces the essential points about the instrument 
and its operation. 
In section 3, FIS-FTS standard data processing is reviewed briefly. 
Section 4 and 5 describe the calibration methods of spectral flux 
and wavenumber scale, respectively. 
The uncertainty of the calibration is summarized in section 6.
In section 7, the reliability of this calibration method is discussed 
by exploring available cross-calibration data.  Some scientific results using
this calibration are discussed by \citet{Yasuda08}, \citet{Okada09} 
and Takahashi et al. (in preparation).

\section{Instrument and Operation Sequence}

\subsection{Instrument}

FIS-FTS has two types of detector arrays covering the wavenumber
range from 55 to 200 cm$^{-1}$. The longer wavelength
array, LW \citep{Doi02}, is a compact stressed Ge:Ga array detector 
and the shorter wavelength array, SW \citep{Fujiwara03}, is a monolithic
Ge:Ga array detector. 
Both the LW and SW arrays have two parts in the arrays, 
corresponding to wider bands (three rows) and narrower bands (two rows), 
respectively, for photometric observations. 
FIS-FTS uses only the wider band part (three rows) of 
each detector array (see figure  3 in \cite{Kawada07}).

\begin{figure*}
\begin{center}
\FigureFile(150mm,150mm){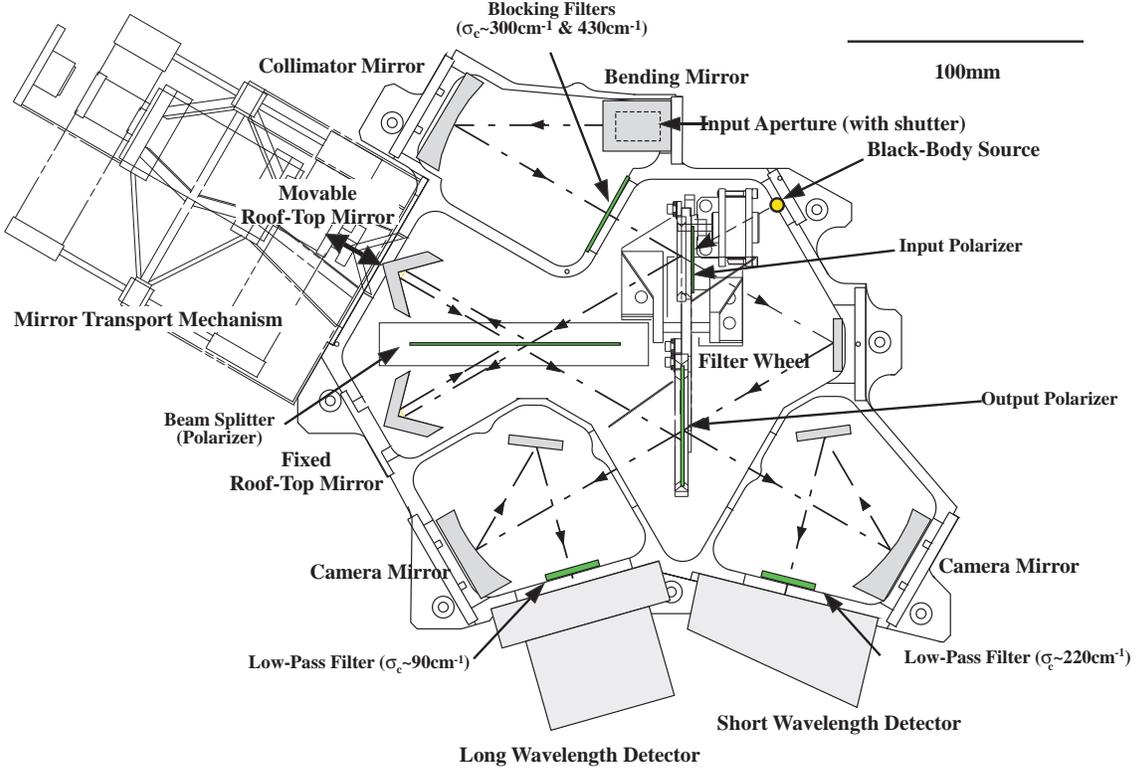}
\end{center}
\caption{Optical design of the FIS-FTS}
\label{fig:OPT}
\end{figure*}

The optical design of the FIS-FTS is described in \citet{Kawada08}. 
Here, we summarize it. The convergent beam
from the telescope is collimated after the bending mirror 
(see figure \ref{fig:OPT}).  The
collimated beam passes through the low-pass filters to 
block photons shorter than 30 \micron. The linearly polarized beam
produced by the input polarizer is divided into two beams by a 
polarizing beam splitter. One of the two beams is reflected by the fixed
roof-top mirror and the other is reflected by the moving roof-top
mirror that changes the optical path difference (OPD) between the two beams. 
After reflection by the roof-top mirrors, the
two beams recombine at the polarizing beam splitter. Finally,
the output polarizer directs the two orthogonal components of the output
beam to the two detector arrays.
FIS-FTS uses a polarizing beam splitter instead of a dichroic beam splitter. 
Therefore, each detector array receives a complementary interference 
signal, i.e. when one array observes a bright central burst (peak), 
the other array observes a dark central burst (valley). 

A small blackbody source is installed on the opposite side of the 
input polarizer for monitoring the responsivity change. 
The temperature of this source can be changed up to 50K. 
The key features of the FIS-FTS are listed in table \ref{tbl:ftsspec}.

\subsection{Operation Sequence}

Astronomical observations using the FIS instrument are performed 
according to the astronomical observation templates (AOTs) of AKARI
 (see \cite{Kawada07}, AKARI FIS Data Users Manual\footnote {AKARI (ASTRO-F) Observers Page - http://www.ir.isas.jaxa.jp/ASTRO-F/Observation/}). 
The AOT labeled FIS03 is dedicated for 
observation using the FIS-FTS. FIS03 is one of  
the pointed observation modes of the satellite and the duration is 
30 minutes including the attitude maneuver operation of the satellite.
During the attitude maneuver from the all-sky survey mode to the 
pointed observation mode, the spectra of the internal blackbody 
source stabilized at 38K are measured with the shutter closed. 
Once the satellite attitude becomes stable, the cold shutter opens and 
the observation starts. After 12 minutes observation, 
the cold shutter is closed again and the maneuver from the pointed 
observation mode to the all-sky survey mode is executed.

FIS-FTS has two spectral resolution operation modes: 
a higher resolution mode (resolution = 0.18cm$^{-1}$) called the ``full-resolution mode''
and a lower resolution mode (resolution = 1.2cm$^{-1}$) called the 
``spectral energy distribution mode'' (SED mode) (table \ref{tbl:ftsspec}). 
Each mode is selected 
by changing the scan length of the moving mirror. 
The mirror scan speeds of both modes are similar.
 Selectable parameters of FIS03 are the spectral 
resolution (full-resolution mode or SED mode), the reset
interval of the detector readout electronics, and 
the on-source position on the detector arrays. We have 
three choices for the last parameter: 
the center of the SW array, the center of the
LW array, or the center of the overlapped area of two arrays.

All measurements of the internal blackbody source are performed with 
the SED mode and a reset interval of 0.5 seconds.  Before opening 
the cold shutter, the parameters are set to the selected values for each
target.  The detector signals and the OPD of the 
interferometer are acquired simultaneously and recorded. 
The data are sampled at a constant time interval. The sampling 
rate of 
the SW array is twice of LW (table \ref{tbl:ftsspec}).

\section{Standard Data Processing}

\begin{figure*}
\begin{center}
\FigureFile(100mm,120mm){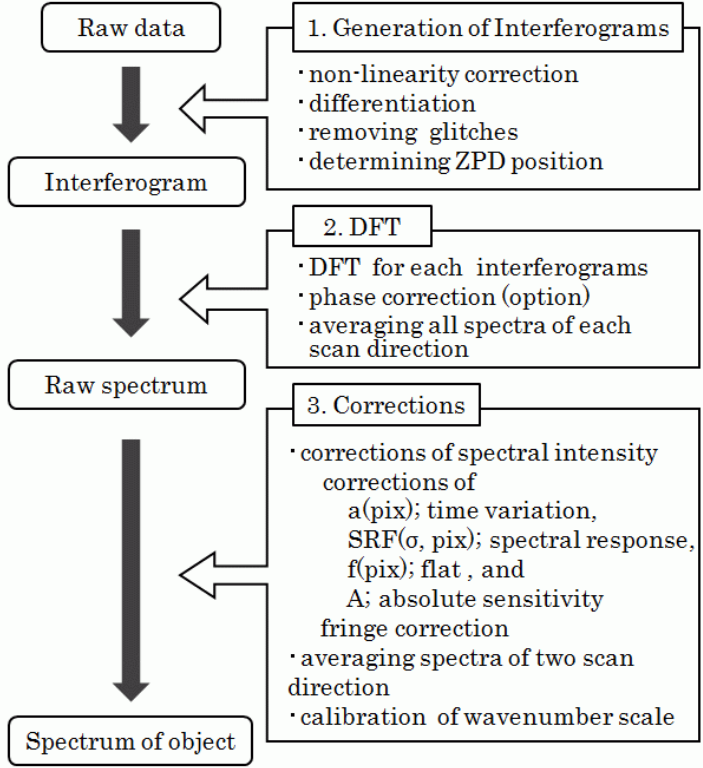}
\end{center}
\caption{Flow chart of the standard data reduction process.}
\label{fig:FLOW}
\end{figure*}

The data analysis flow of FIS-FTS taken with FIS03 is as follows:
\begin{enumerate}
\item Generation of interferograms by combining the detector signals 
with the OPD
\item Calculation of spectra by discrete Fourier transformation
of the interferograms
\item Corrections to obtain the spectra of the objects
\end{enumerate}
This sequence is shown in figure \ref{fig:FLOW}. 
These procedures are involved in the FIS-FTS data processing pipeline 
(AKARI FTS Toolkit Manual\footnote {AKARI (ASTRO-F) Observers Page - 
http://www.ir.isas.jaxa.jp/ASTRO-F/Observation/}) as the standard
process.
Each step is described separately in the following subsections.

\subsection{Generation of Interferograms}

The far-infrared detector signals are read out using the integration
amplifier circuit. After correction for the non-linearity of the 
readout electronics (see AKARI FIS Data Users Manual), 
the data are differentiated one by one with 
the time-sequential signals sampled with a constant interval. 
Glitches caused by charged particles striking the detector or 
electronics (\cite{Suzuki08}) are removed from the signals at this stage.  
The sampling interval in the OPD depends 
on the scanning speed of the moving mirror. The speed of the 
moving mirror is approximately 0.07--0.08 cm s$^{-1}$ on average; 
however, it is not constant. 
Near the end of the mirror travel, the speed decreases 
by 10\% due to 
varying leaf spring tension of the mirror transport mechanism. 
Moreover, a 15 Hz modulation has been found in the mirror speed.
It can be attributed to the mechanical 
interference from mechanical coolers in the AKARI satellite. 
The amplitude of this velocity modulation is 15\% of the average 
velocity (see \cite{Kawada08}).
  As the interferograms are sampled with 
a frequency five times higher than the Nyquist criterion, 
we have sufficient oversampling to study the effect of nonuniform
sampling and find that it is negligible.

\begin{figure*}
\begin{center}
\FigureFile(150mm,150mm){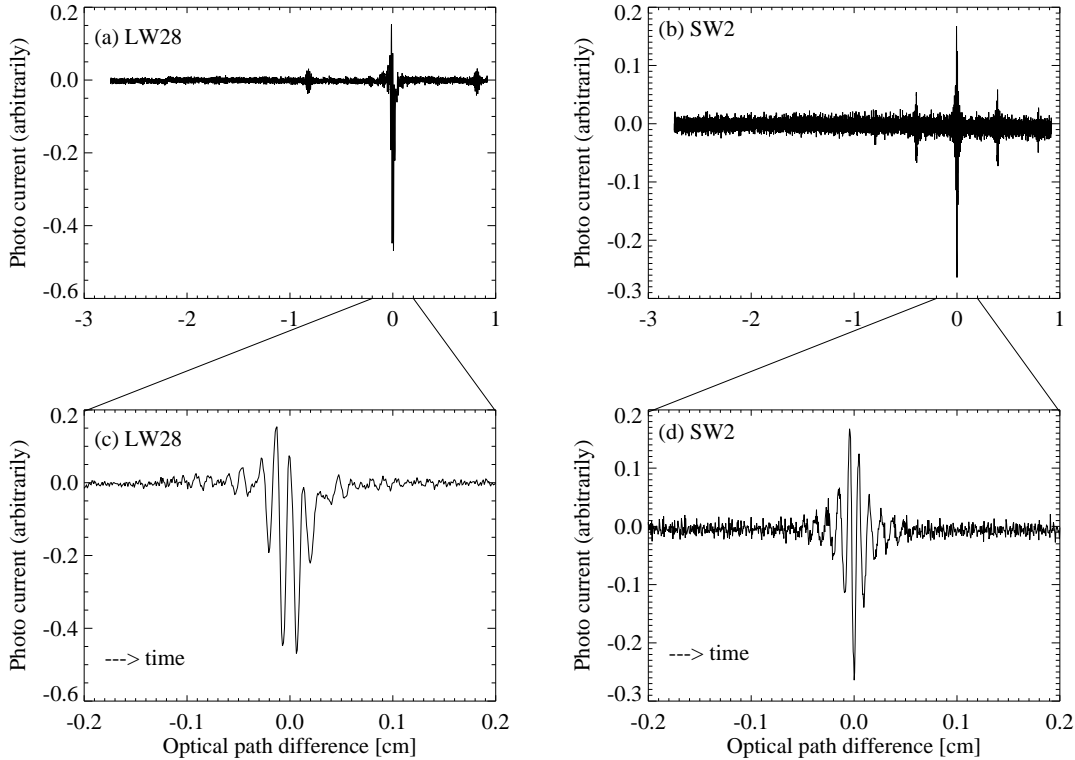}
\end{center}
\caption{Examples of interferograms measured by the FIS-FTS. Panel (a) and (b)
    are full range interferograms taken in the full-resolution mode
    of LW (LW28) and SW (SW2),
    respectively. Symmetric sub-structures can be seen along the optical
    path difference positions in both panels.   Panels (c) and (d)
    are close-up views around the ZPD positions of panels (a) and
    (b), respectively. 
    Asymmetries due to the transient response of 
    the detectors are clearly seen.
}
\label{fig:IF}
\end{figure*}

FIS-FTS has a displacement sensor for the moving mirror that is used to measure 
the relative OPD; however, it lacks a sensor for the 
detection of the zero optical path difference (ZPD) position. 
The ZPD position is determined from the interferograms 
by minimizing the integral of the imaginary part of the spectra 
as a function of the origin of the Fourier transformation. 
Some examples of
interferograms produced by the data processing pipeline are shown in
figure \ref{fig:IF}. At this stage, the OPDs for the thermal 
contraction of the optical scale of the displacement sensor and the
angle between the scale and the optical axis of each pixel have not 
been corrected.
These corrections are easy and applied in the spectral domain (section 5). 

\subsection{Discrete Fourier Transformation (DFT)}

The DFT method is adopted in the FIS-FTS data processing pipeline, 
which calculate a numerical integration of the associated Fourier 
integrals, instead of the FFT algorithm to derive better results 
for non-uniform sampling data.  The spectra obtained at this step 
(hereafter, raw spectra) are calculated for each scan.

In a typical observing sequence, 11 forward and backward 
scans of the internal blackbody
source are recorded, followed by 7 or 30 
forward and backward scans of the astronomical
source in the full-resolution mode or the SED mode, respectively.
In the SED mode, the interferograms are double-sided with a maximum
OPD of $\pm$0.42 cm. In the full-resolution mode, 
the interferogram is asymmetric, and the maximum OPD 
of one side is expanded to 2.7 cm and the other side to 0.9 cm. 
The phase information is derived from short, doubled-sided 
interferograms and applied to single-sided full span data to derive full 
resolution spectra.
Since the interferograms are asymmetric in opposite directions between 
the forward and backward scan direction, spectra from two directions 
are averaged and calibrated individually.

\subsection{Corrections}
The raw spectrum corresponds to a product of the spectrum of the
source and the total spectral response of the FIS-FTS which includes 
the spectral response of the detectors, the efficiency of optical filters, and 
the modulation efficiency of the interferometer.
To derive the spectrum of the source from the raw spectrum, 
flat correction, correction of time variation of the detector
responsivity, spectral response calibration and absolute flux calibration are required.
In the case of line intensity measurements, the fringe pattern remaining 
in the spectrum must be removed.
Finally, the source spectrum is obtained by averaging the 
calibrated spectra of both scan directions with the 
wavenumber scale correction.
The derivation of these correction factors and functions 
is described in sections 4 and 5.

The FIS-FTS detectors reveal the transient response 
(\cite{Hiromoto99,Kaneda02}) that affects the accuracy of the 
measured spectra 
through the distortion of the interferograms. 
This distortion is clearly 
observed around the zero path position (figure \ref{fig:IF}), 
and varies from pixel to pixel, especially for LW. 
We do not attempt to correct the transient effect 
in time domain in this paper because of the complexity of the non-linear 
behavior of the photoconductive detectors. The shape of Fourier transformed 
spectra, within the effective wavelength range, are nearly free from the 
distortion of interferograms (see \cite{Kawada08}). Furthermore, since 
the same procedures are applied to both calibration and target 
source data, the effect of the transient response can be canceled as the 
zeroth-order approximation.

\section{Correction of Spectrum}

The object spectrum can be derived by the following equation as a
function of both wavenumber ($\sigma$) and pixel ID (pix):

\begin{equation}
 I(\sigma, pix)=i(\sigma,  pix) \times a(pix) \times A \times f(pix) \times \frac{1}{SRF(\sigma, pix)}, \label{eq:1}
\end{equation}

\noindent where

 $I(\sigma,pix)$ : the object spectrum at certain pixel [Wm$^{-2}$Hz$^{-1}$sr$^{-1}$],

 $i(\sigma,pix)$ : the raw spectrum [Vsec$^{-1}$Hz$^{-1}$],

 $a(pix)$ : the correction factor 
of the time variation of the detector responsivity at each observation,

 $A$ : the absolute flux calibration factor [W m$^{-2}$ sr$^{-1}$V$^{-1}$sec],

 $f(pix)$ : the flat correction factor, and

 $SRF(\sigma,pix)$ : the relative spectral response function 
which includes the spectral response of both the detector and the optics.

\noindent $A$ is a common value for all pixels while other factors $a$, $f$, and $SRF$ are 
determined for each pixel independently.

We describe briefly how to derive the factors.
$A$ is derived from comparison of observed spectra of solar system objects 
with their models. 
$a(pix)$ is determined by measuring the internal 
blackbody source at every observation.
$f(pix)$ is derived from the measurements of the aperture lid of the telescope
referring to the internal blackbody source. 
The telescope aperture lid was measured before launch.
To determine $SRF(\sigma, pix)$ of all pixels, many astronomical 
sources were used, whose spectra had been measured by the ISO/LWS or are
well determined by models.
The data used in the calibration schema are described in 4.1. The
details of calibrations and fringe correction are explained in 4.2 -- 4.6.

\subsection{Calibration Sources}

\begin{table*}
  \caption{Calibration sources.}\label{tbl:cal_obs_list}
  \begin{center}
    \begin{tabular}{lllll}
      \hline
      name & mode & no. of pointed obs. & reference & flux (@100\micron)\\
      \hline
      point source\\
      \hline
      Uranus & SED & 1 (2006/06/05 20:38) & model\footnotemark[$*$]  & 860Jy\\
      Neptune & SED & 1 (2007/05/12 22:19)  & model\footnotemark[$*$] & 340Jy\\
      Ceres & SED & 1 (2007/08/12 03:59)& model\footnotemark[$\dagger$] & 160Jy\\
      \hline
      extended source\\
      \hline
      \multirow{2}{*}{M82} & SED & 2 (2006/04/19) & \multirow{2}{*}{ISO/LWS\footnotemark[$\ddagger$]} & \multirow{2}{*}{$\sim$10GJy sr$^{-1}$}\\
       & full-res. & 7 (2006/04/19, 2006/10/22) & &\\
      Galactic center & full-res. & 5 (2006/09/19, 2007/03/17-18) & ISO/LWS\footnotemark[$\ddagger$] & $\sim$ 35GJy sr$^{-1}$\\
      M20 & SED & 1 (2006/09/23)& ISO/LWS\footnotemark[$\ddagger$] & 3--5 GJy sr$^{-1}$\\
      M17 & full-res. & 1 (2006/09/27)& ISO/LWS\footnotemark[$\ddagger$] & 5--37 GJy sr$^{-1}$\\
      $\eta$ Carinae & full-res. & 6 (2007/01/12-13, 2007/07/14) & ISO/LWS\footnotemark[$\ddagger$] & 2--8 GJy sr$^{-1}$\\
     \hline
     \multicolumn{4}{@{}l@{}}{\hbox to 0pt{\parbox{120mm}{\footnotesize
	 
	   \footnotemark[$*$]M\"{u}ller \& Lagerros (1998, 2002),
	   \footnotemark[$\dagger$]Moreno (1998), 
	   \footnotemark[$\ddagger$]Archival Data
	 }\hss}}
    \end{tabular}
  \end{center}
\end{table*}

\subsubsection{Astronomical Objects}

Table~\ref{tbl:cal_obs_list} lists all the astronomical sources used
for the calibration of the FIS-FTS. 
They are bright point sources whose fluxes are 200--900 Jy at 
100 \micron\, or  bright extended sources whose intensities are larger than a few GJy sr$^{-1}$.
They are used to obtain the relative response function of $SRF(\sigma, pix)$ while 
some of the solar system objects are used for the absolute calibration 
of the intensity. The flux of the solar system objects changes due to
their rotations and the observing geometries. 
The flux at the time of each observation was thus calculated using
the appropriate models (\cite{Muller98}, \yearcite{Muller02}; \cite{Moreno98}). 
The uncertainties of the model spectra are smaller than about 
5\%, 20\%, and 10\% in the wavelength range of
the FIS-FTS for Uranus, Neptune and Ceres, respectively. 
Uranus and Neptune were used to derive the absolute
calibration factor of the FIS-FTS. These sources are frequently used as
primary calibrators at far-infrared wavelengths, and Uranus was the
principle calibration source for the ISO/LWS \citep{Gry03}. 
Because the apparent
diameters of these objects are smaller than five arcseconds, they can be
considered as point sources for the FIS-FTS. 

The extended sources listed in table 2 were observed by both the ISO/LWS and FIS-FTS. 
The calibration using these extended sources is more efficient compared to 
the solar system objects because many FIS-FTS pixels can be calibrated simultaneously.
However, the uncertainty in the absolute intensity is larger than that 
derived with solar system objects, because the absolute calibration of 
the ISO/LWS itself has rather large uncertainty. Thus these sources are only 
used to derive $SRF(\sigma,pix)$.

The ISO/LWS spectroscopic observations with LWS01 (full-grating scan
mode) overlap the wavelength range covered with the FIS-FTS. Using
the Off-Line Processing (OLP) version 10.1 data obtained from the
ISO Archival Data Center, the LWS spectra are processed (removing
glitches, averaging, defringing, and shifting to match the vertical
level of the data of different detectors) using the ISO Spectral
Analysis Package (ISAP\footnote{The ISO Spectral Analysis Package
(ISAP) is a joint development by the LWS and SWS Instrument Teams
and Data Centers. Contributing institutes are CESR, IAS, IPAC, MPE,
RAL, and SRON.}).

\subsubsection{The Internal Blackbody Source}

The internal blackbody source is a bright diffuse source that
irradiates both of the two detector arrays and the signal-to-noise
(S/N) ratio is high for almost all the pixels. 
However, because of its location in the instrument (figure \ref{fig:OPT}), 
the polarity of the
interferogram of the internal blackbody source is opposite to that
of astronomical sources, 
which can induce a different distortion on the interferograms between 
the internal source and astronomical sources.
Therefore, {\it SRF} cannot be made directly from the measurement
of the internal blackbody source. On the other hand, since the internal
blackbody source is measured at every observation, these
measurements can be used 
to compare the on-orbit performance of the instrument with that in the
laboratory before launch, and
correct the time variation of the detector responsivity.

\subsubsection{Aperture Lid}

The aluminum aperture lid 
on top of the telescope baffle covers the entire field 
of the telescope in order to block radiation from the hot surface 
before the operation.  The aperture lid could be a flat source for the 
array detectors. However, its extremely high temperature saturate the detectors. 
At the ground test of the satellite, the 
special aperture lid was used for the evaluation of instruments 
and was cooled by an extra liquid helium bath. 
The temperature of the cooled aperture lid is 
about 40 K, and all walls in the 
telescope cavity should be colder than the lid. The cooled aperture lid can be 
assumed to be a flat source because all the detector pixels see almost
the same area of the lid and the scattered light from the outer area
is negligible. 
Therefore, the measurement of the cooled aperture lid 
can be used as a reference for the flat-field correction.

\subsection{Correction for Time Variation}

The detector responsivity changes mainly 
due to the radiation environment 
in space. This responsivity change can be 
traced by measurements of the internal blackbody source at 
every observation.
Both the measured spectral shape and the intensity are stable 
within 10\% throughout the entire 
observation period of the FIS-FTS, except only for extremely high sensitivity
cases that are activated by high radiation background.
After scaling the detector responsivity according to the integrated signal 
of the internal blackbody source,
the remaining variation in the spectral shapes decreases to 
3--5\%. 
The correction factor $a(pix)$ is given by the ratio of the integrated 
signal of the internal blackbody source at each observation to the averaged 
value of observations.

\subsection{ Spectral Response }

\begin{figure*}
\begin{center}
\FigureFile(80mm,80mm,angle=90){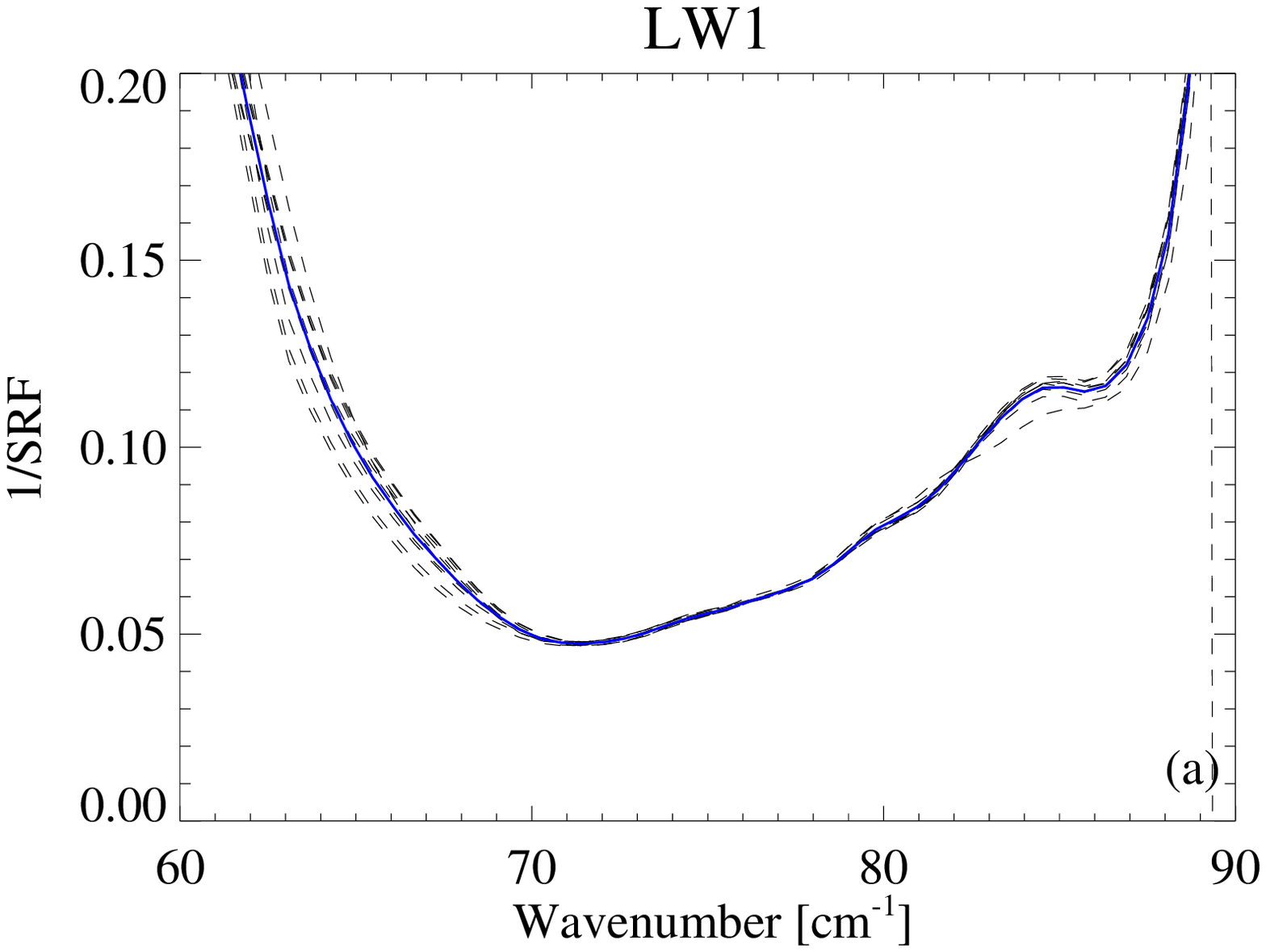}
\FigureFile(80mm,80mm){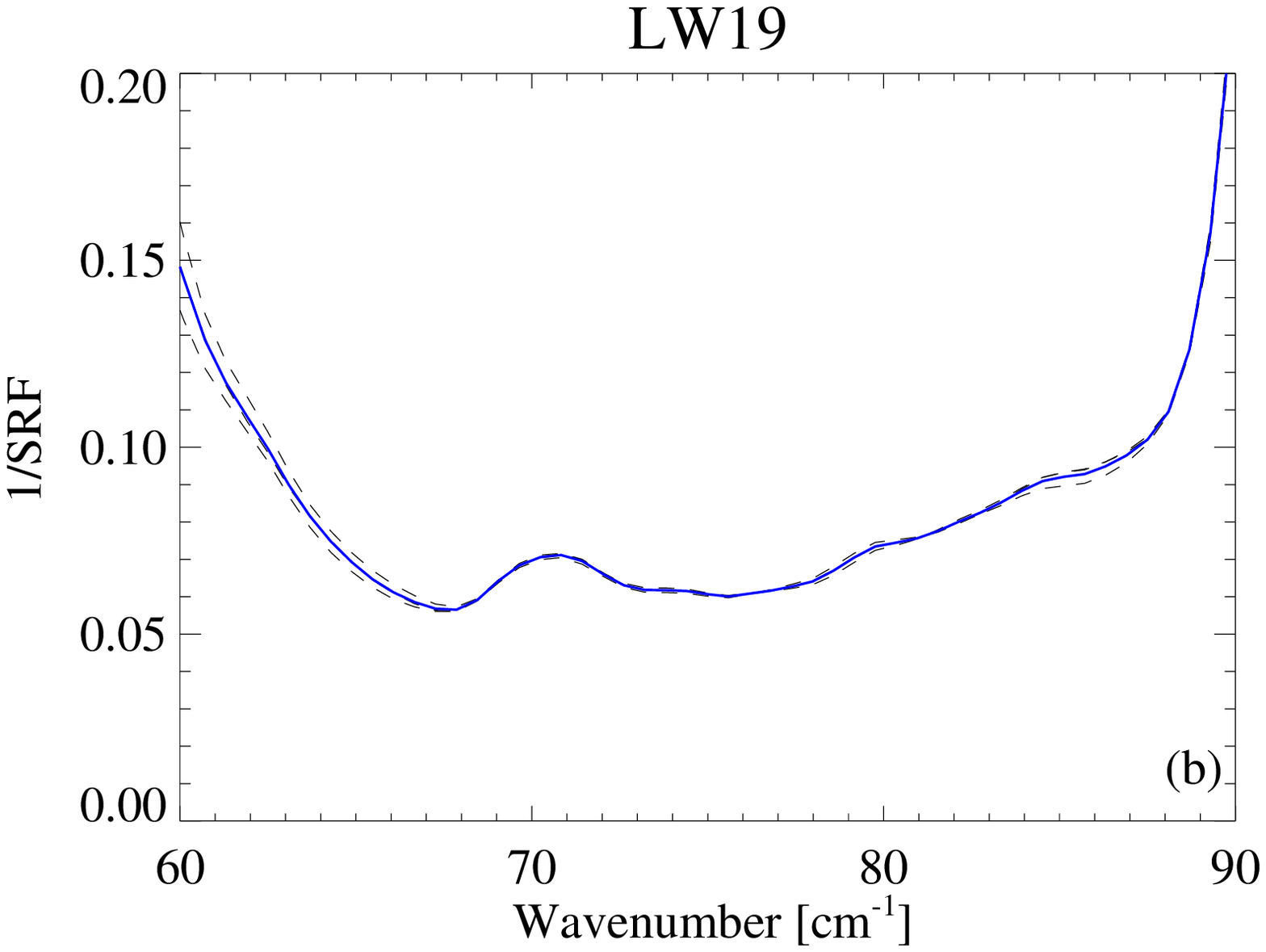}
\FigureFile(80mm,80mm){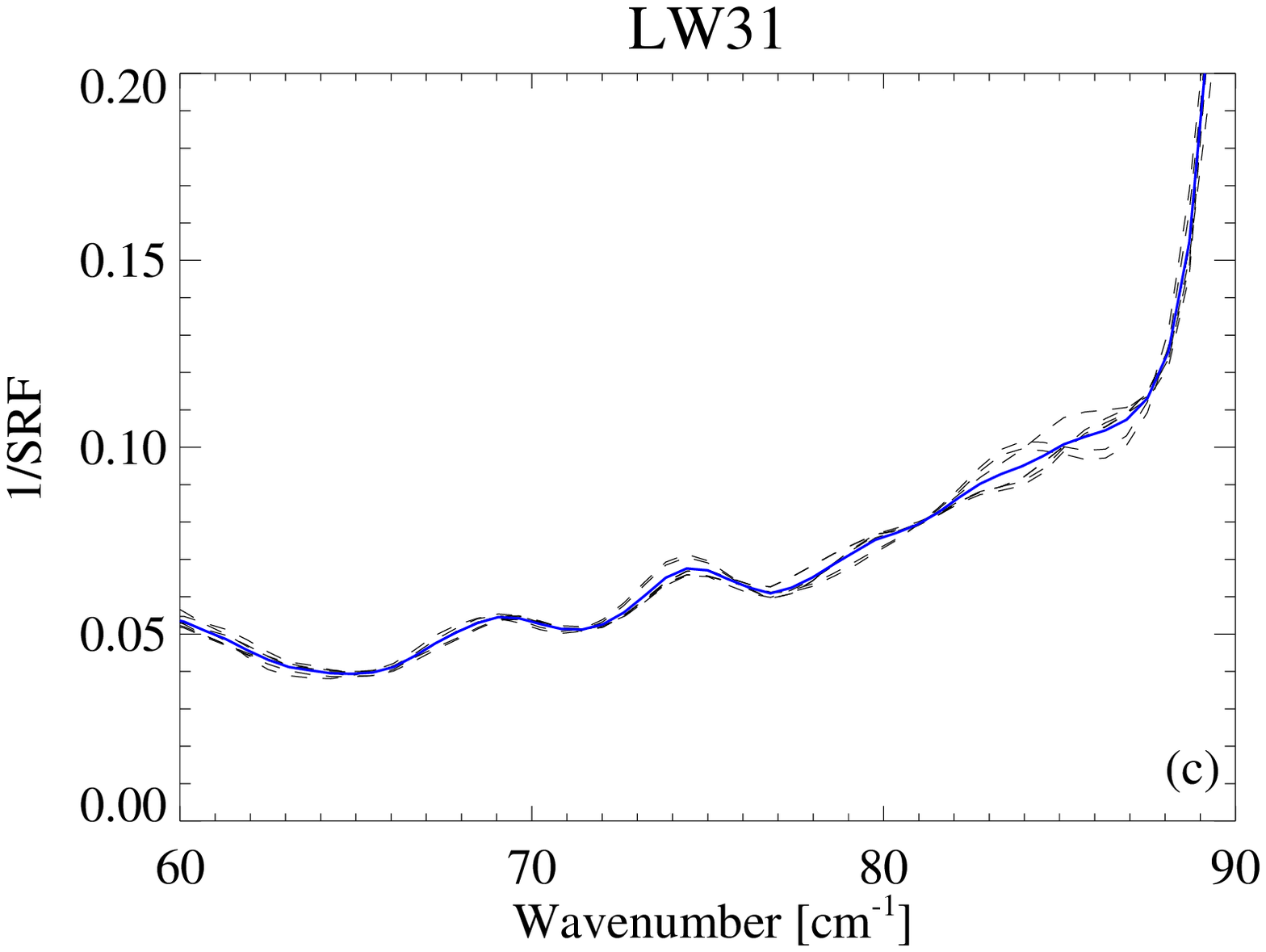}
\FigureFile(80mm,80mm){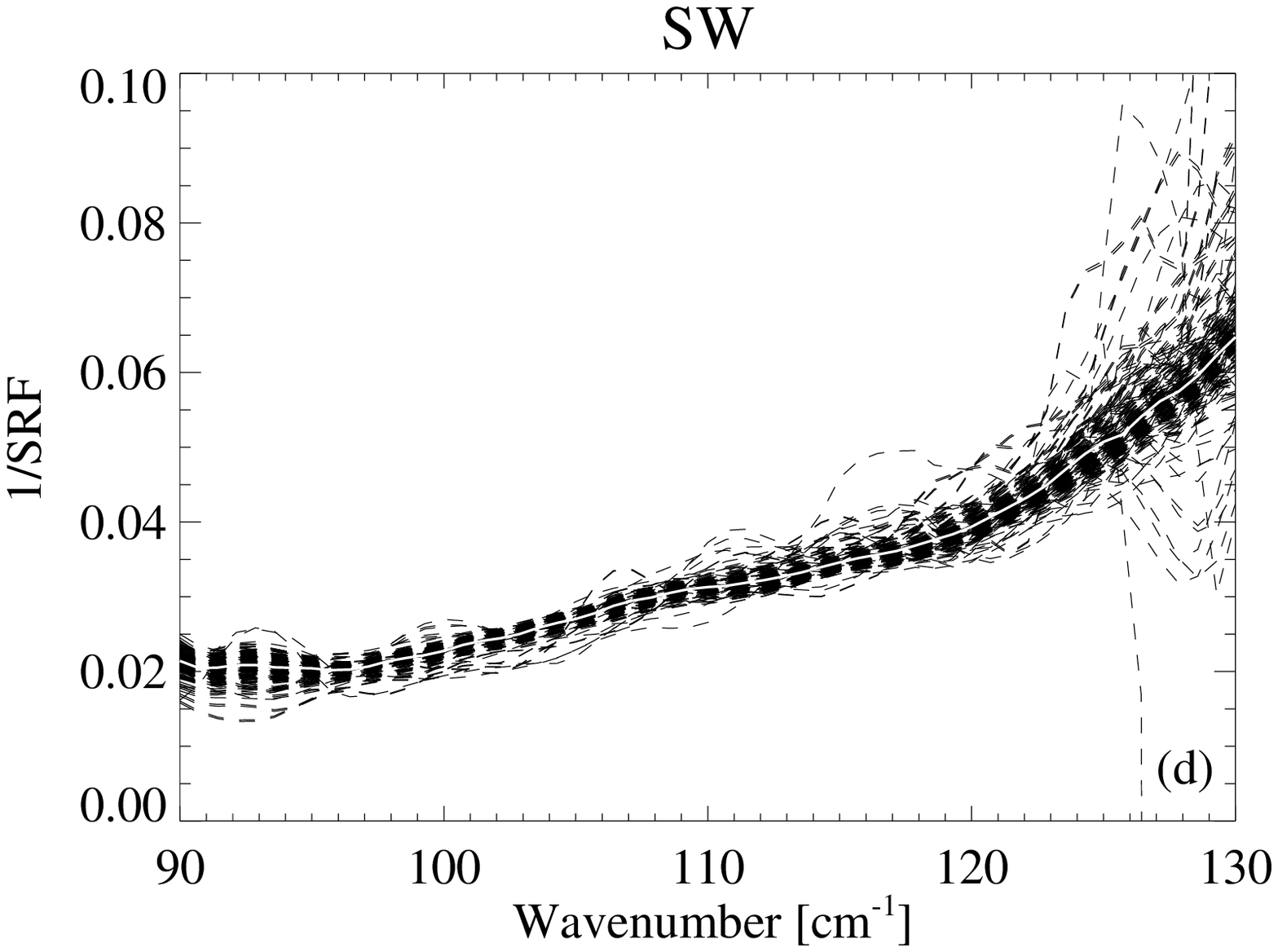}
\end{center}
\caption{
Examples of $SRF$s as shown by the inverse of the $SRF$ (thick solid line)
        with all $SRF$s before averaging. 
        The LW pixels have a wide variety in shape as shown in 
        panels (a) - (c).  On the other 
	hand, the SW pixels have similar shapes as shown in panel (d).  
        The $SRF$ averaged over all pixels has 
	been adopted as the common one for the SW array.}
\label{fig:ref_div_akari}
\end{figure*}

The spectral response function, $SRF(\sigma,pix)$, includes all the 
wavenumber($\sigma$) dependence in the flux calibration
(Eq. \ref{eq:1}) and is 
obtained from the ratio of the reference spectrum, $I(\sigma, pix)$, to the 
measured raw spectrum, $i(\sigma, pix)$, after appropriate normalization.
The reference spectra are the ones from the models or 
ISO/LWS observations (table~\ref{tbl:cal_obs_list}).
To avoid the fringe structure in spectra described below, 
short span interferograms are transformed to make raw spectra.
 After that, the emission lines are masked and smoothed by 
averaging 5 points to get spectral resolution of about 6.5 cm$^{-1}$ 
with better S/N ratio.
Thus, $SRF(\sigma, pix)$s obtained are shown in figure~\ref{fig:ref_div_akari} as the
inverse of $SRF(\sigma, pix)$. The pixel-to-pixel variation in the SW
array is small 
enough to adopt the average as a single common $SRF$ for all pixels 
as shown in panel (d). 
In contrast, for the LW array, $SRF(\sigma, pix)$ of each pixel is 
determined one by one, since the 
pixel-to-pixel variation of $SRF(\sigma, pix)$ is large.

The scatter of all the $SRF$s that are used to derive the averaged 
$SRF$ described above is 5--10\% for the SW array (Fig. 4d) 
and within 5\% for the LW array in 70--85 cm$^{-1}$ (e.g., Fig. 4a-c), 
although it is several to 20\% depending on pixels in the 
60--70 cm$^{-1}$ region, where the difference 
in responsivity of individual pixels is relatively large.

\subsection{Flat Correction}

For flat correction, we use the measurement of the cooled 
aperture lid of the telescope (section 4.1.3) 
and the internal blackbody source. 
The measurement of the aperture lid was performed 
on the ground before launch, when the lid was considered to be a uniform radiator. 
The relative sensitivity between pixels could change after the launch as the
result of changes in the effective detector bias voltage on each pixel and 
the radiation effects.
Therefore, we cannot apply the flat correction factor derived from the 
measurement of the aperture lid directly to the observational data.
We take spectra of the internal blackbody source both on the ground and in flight. 
These spectra are not completely flat but their illumination pattern is quite stable. 
We determine the flat correction factor in flight from the combination of the 
measurements of the aperture lid and the internal blackbody source.

\begin{figure*}
\begin{center}
\FigureFile(80mm,80mm){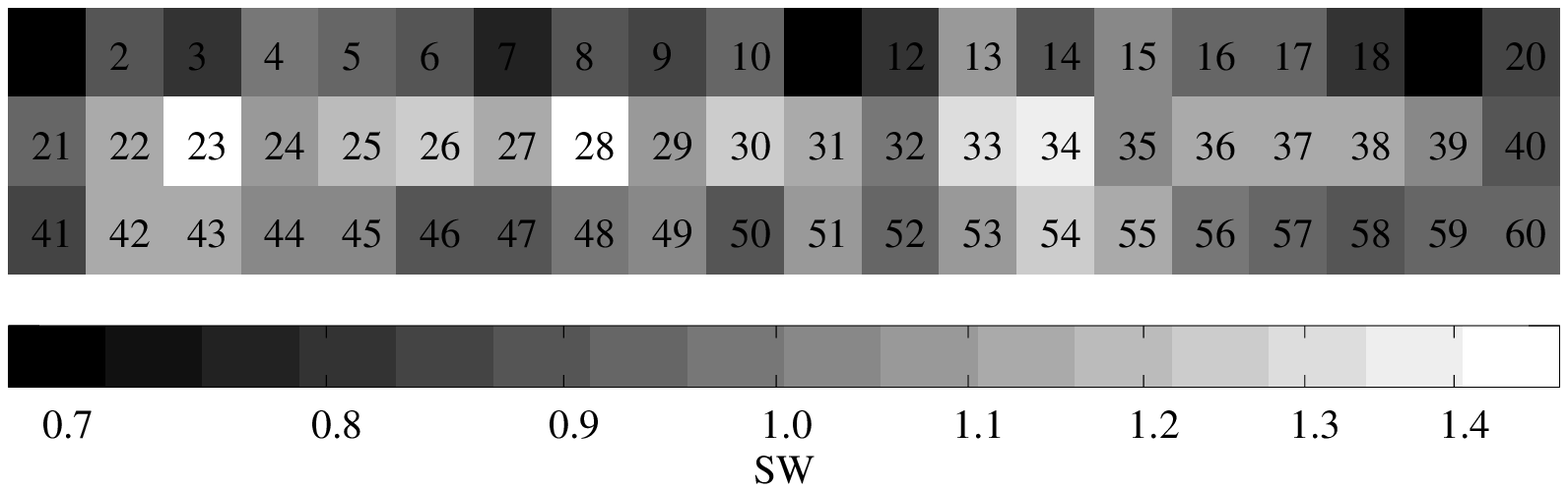}
\FigureFile(80mm,80mm){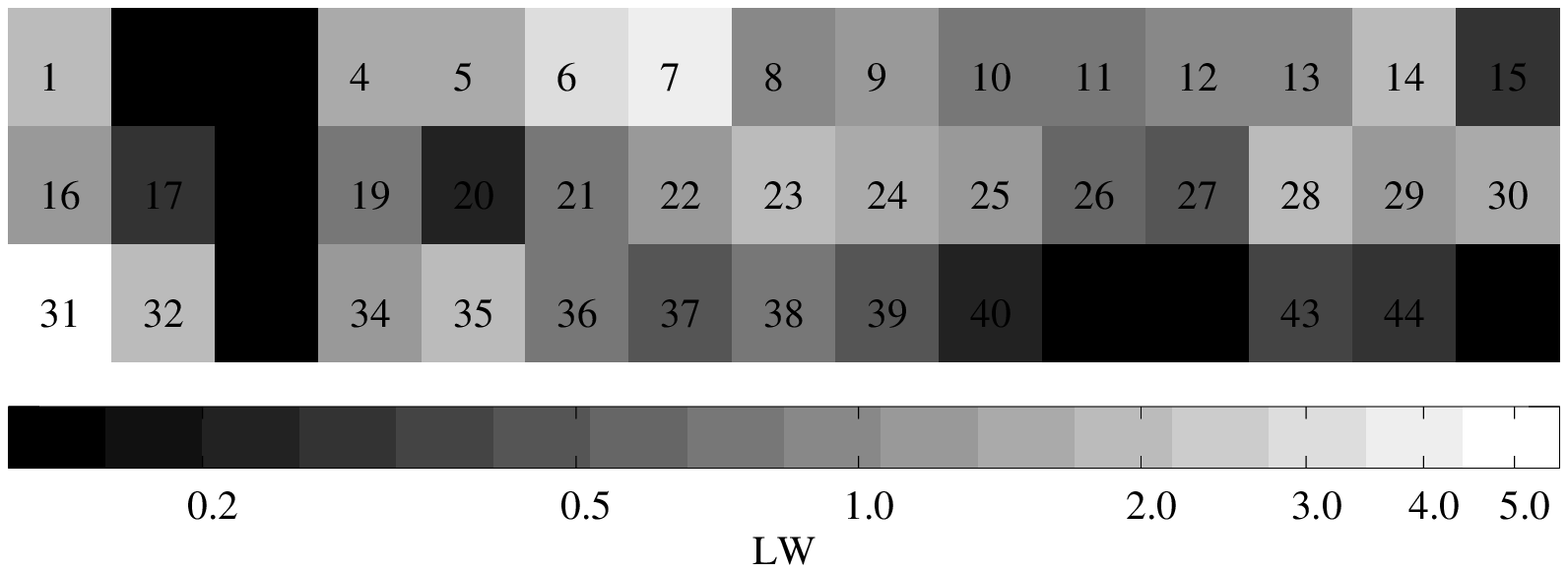}
\end{center}
\caption{
        Flat correction factors of SW (left) and LW (right).  
        The gray scale indicates the relative responsivity to the median values.  
        Each pixel is labeled with a serial number.
        The pixels of SW1, SW11, LW2, LW3, LW18, LW33, and LW42 
	did not work in space.}\label{fig:flat}
\end{figure*}

We calculate the integrated power of derived spectra within the most 
sensitive wavenumber range (70--85 cm$^{-1}$ for LW 
and 90--125 cm$^{-1}$ for SW)
for both the aperture lid ($P_{lid,0}$) and the internal
blackbody source, measured on the ground ($P_{BB,0}$) and in
orbit ($P_{BB,1}$). 
Here, $P_{BB,1}$ is calculated for the averaged spectrum of some 
of the blackbody source measurement in orbit, 
and $P_{BB,0}$ is calculated for the spectrum measured 
during the same experiment of the aperture lid measurement.  
The flat correction factor, $f(pix)$, is determined as
\begin{equation}
	f(pix) = P_{lid,0}(pix) \times \frac{P_{BB,1}(pix)}{P_{BB,0}(pix)}. \label{eq:3}
\end{equation}
The first term on the right-hand represents the flat on the ground, 
and the second term is the correction for the sensitivity change in orbit.
Figure \ref{fig:flat} shows the distribution of f(pix) for the two detector arrays. 
A large variation, over 10 times the responsivity in the pixels
of the LW array, is observed. 
This variation is caused by non-uniformities in 
the effective detector bias voltage and in the effective spectral
response (\cite{Kawada07}).
Assuming that the aperture lid illuminates the array pixels with complete
uniformity, the accuracy of the flat correction is around 10\%, which 
arises from a relative calibration uncertainty due to the 
internal blackbody source.

\subsection{Absolute Calibration}

For determination of the absolute flux calibration factor {\it A}, 
we use the observations of Uranus and Neptune since their fluxes are 
accurately determined by the models (section 4.1.1) and their S/Ns are high 
enough. Each of them is observed only once at the appropriate position, 
and, thus, 
only two pixels of each detector can be calibrated directly.  
The on-pixel flux is 
calculated with the source position on the pixel and the point spread 
function of the FIS-FTS.  
The position uncertainty of the source on the pixel results in the 
uncertainty of $+20\%/-50\%$ in the on-pixel flux and, thus, of the factor A.
This uncertainty remains 
the systematic error of the absolute calibration for all observations.

\subsection{Fringe Correction}

The interferograms of the FIS-FTS exhibit 
clear sub-structures that are symmetric about the ZPD 
(see figure \ref{fig:IF}). 
This pattern is called, ``channel fringe''. Such 
features are common in Fourier spectroscopy,
particularly at longer wavelengths, and associated with resonant
optical cavities within the interferometer \citep{Naylor88}. There
are two different causes in the FIS-FTS: 
(1) multiple-beam interference between the two blocking filters, and 
(2) multiple-beam interference in the Ge:Ga detector substrate. 
Whereas the former could be seen both in LW and SW,
the latter could be seen only in the SW spectra.  
Since the causes of the channel fringes are identified
with well-defined physical properties, the features are expected to
be stable and reproducible, allowing their correction using
physical models.

\begin{figure*}
\begin{center}
\FigureFile(80mm,80mm){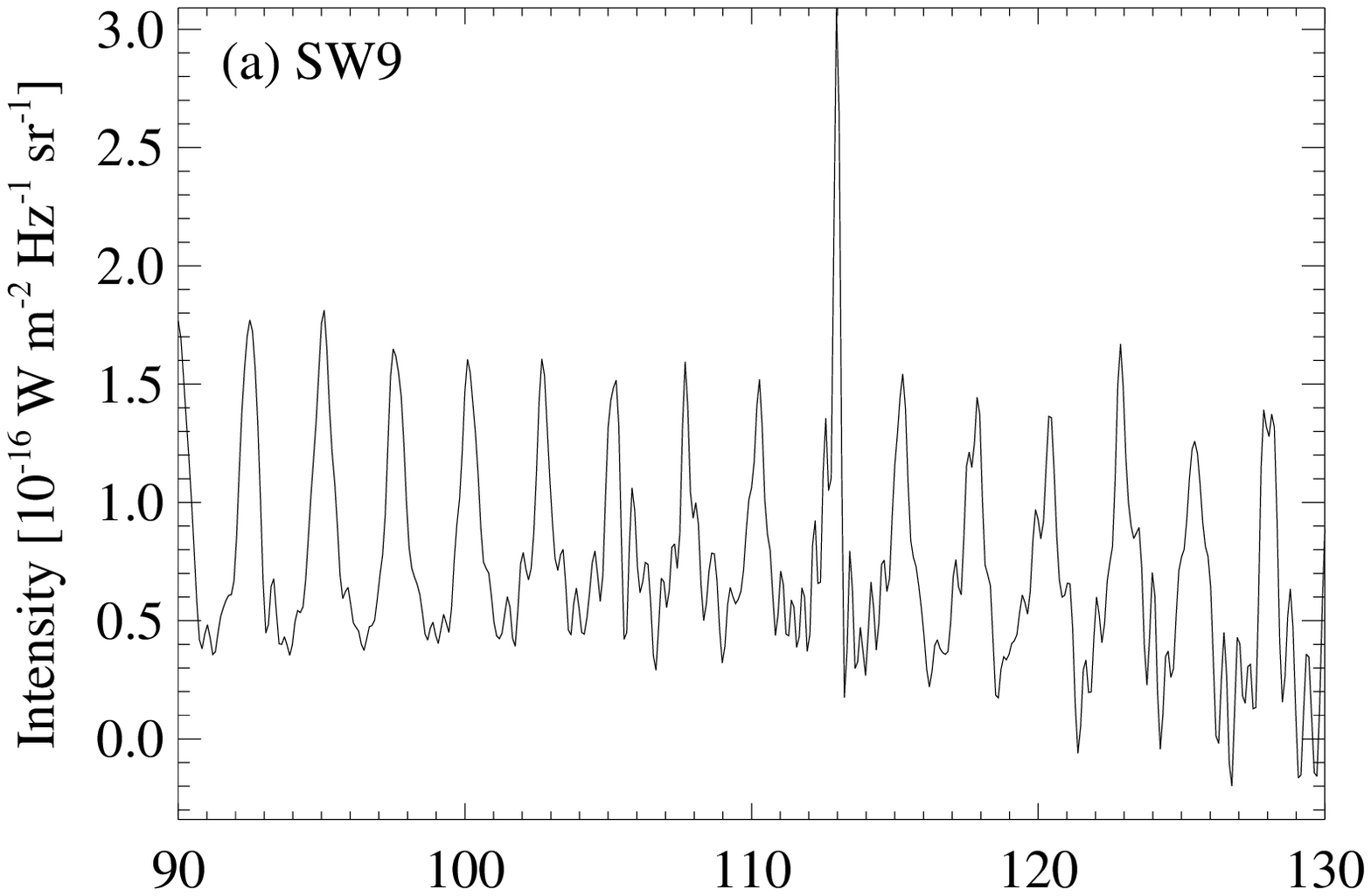} 
\FigureFile(80mm,80mm){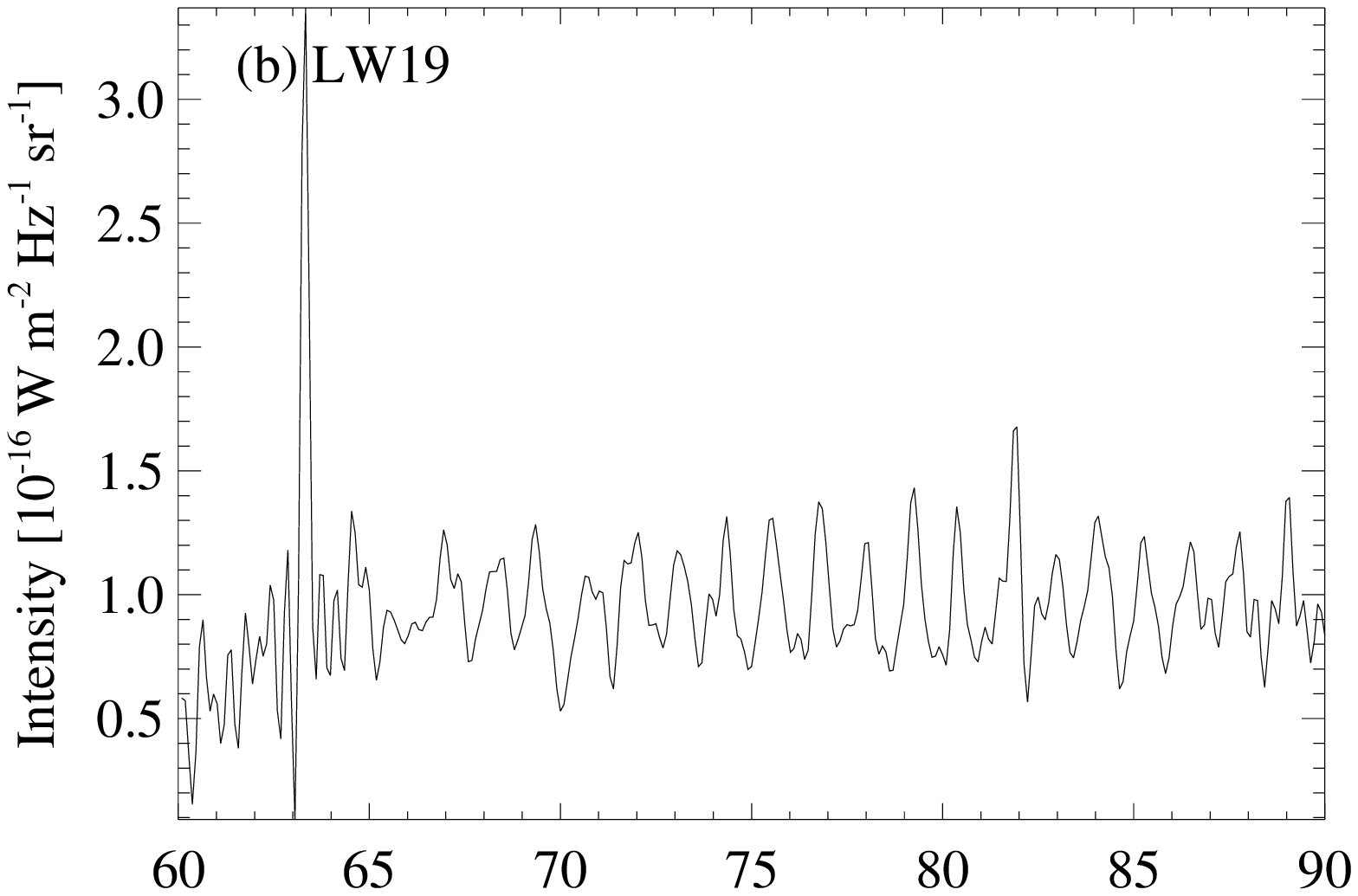}
\FigureFile(80mm,80mm){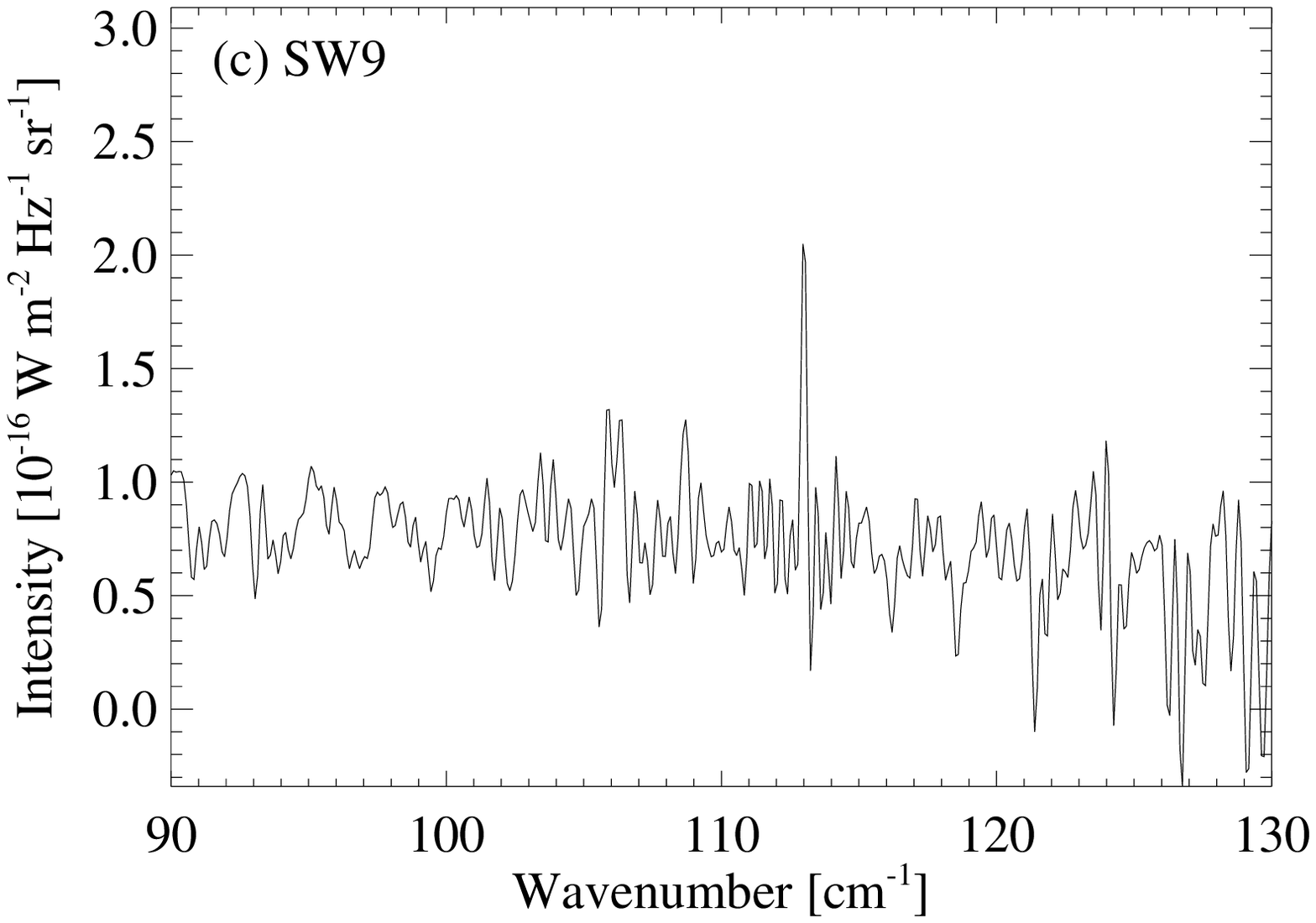}
\FigureFile(80mm,80mm){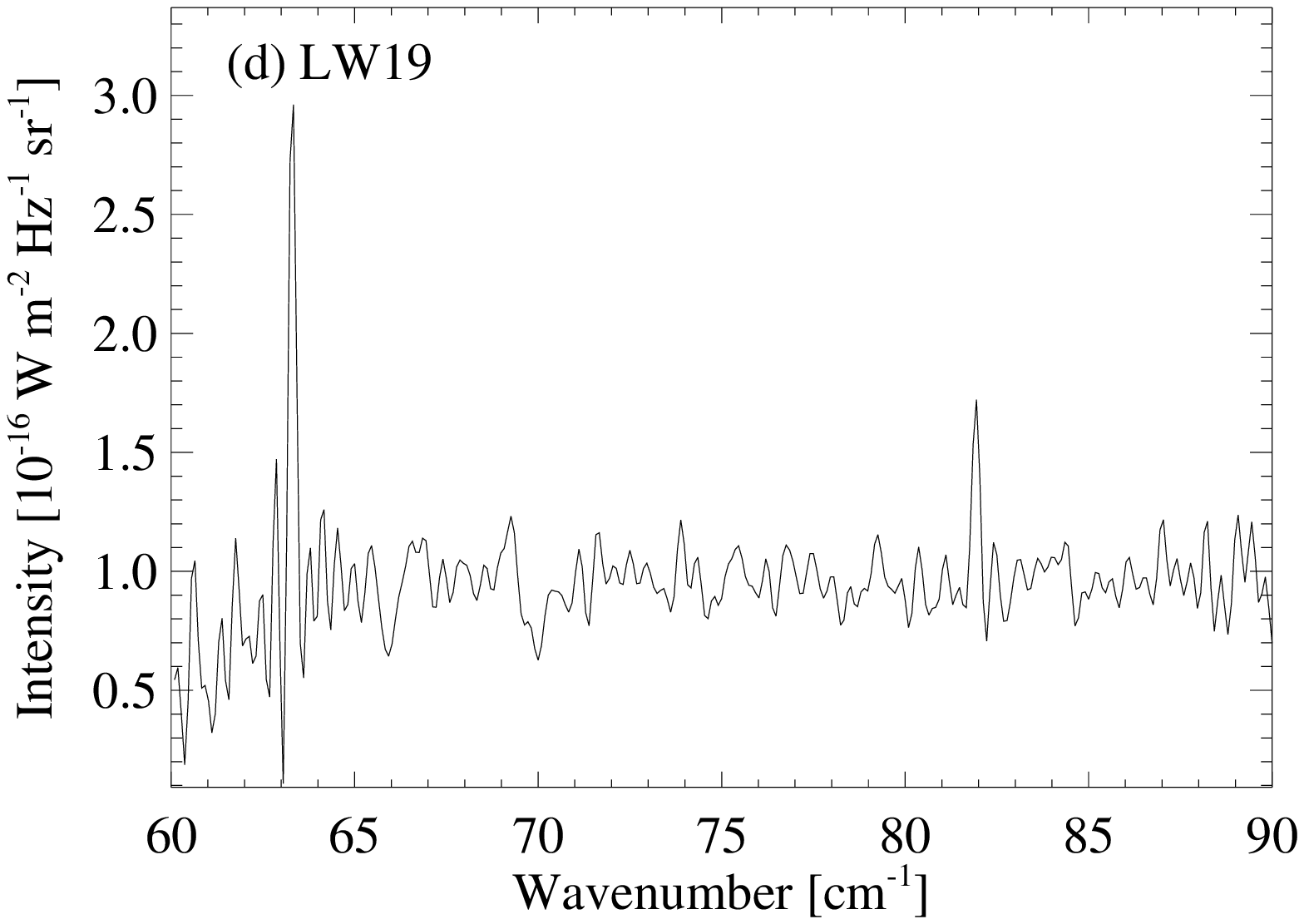}
\end{center}
\caption{Examples of the full-resolution spectra of G333.6--0.2 including 
        emission lines. Panels (a) and (b) show the spectra
	of SW9 and LW19 before defringing, respectively.
        The fringe pattern is clearly seen in both spectra. The periodicity 
        is different in each case as described in the text.  
        After the defringing procedure, 
	the emission lines appear clearly as shown in panels (c) and (d), 
        processed from (a) and (b), respectively. The fitting ranges for 
        the local fringe are 8 cm$^{-1}$ in both sides around 
	[O\,{\scriptsize III}] (113 cm$^{-1}$) in the case of panel (c)
	and 4 cm$^{-1}$ in both sides around [N\,{\scriptsize II}] 
	(82 cm$^{-1}$) in the case of panel (d).
	Note that there can be recognized a pattern of the sinc
        function at every line because the apodization is not applied.
}
\label{fig:defringe}
\end{figure*}

Upon Fourier transformation the channel fringes
produce an oscillatory component in the spectrum (figure
\ref{fig:defringe}). Since $SRF$s are determined by using low
resolution spectra ($ \Delta\sigma = $ 6--7 cm$^{-1}$), $SRF$s 
do not correct the fringe pattern that 
appeared in the higher spectral resolution spectra.
For emission line extraction, the correction of the fringe pattern 
is essential.
A theoretical model of Airy's formula (\cite{Born75}) for the 
resonant optical cavity is applied. 
The parameters of Airy's formula are determined by fitting 
to several full-resolution spectra of the Galactic Center after exclusion of 
strong emission lines.
Since the reflectivity depends on wavenumber, 
fitting by Airy's formula is applied to a narrow wavenumber
range around the target emission line.  Two examples of the
defringing spectra are shown in figure \ref{fig:defringe}.
After the defringing procedure, the emission lines appear clearly 
as shown in panels (c) and (d).

\section{Scale Correction in Wavenumber}

Through the preceding section, the raw values of the 
optical scale of the displacement sensor for the moving mirror (section 3.1)
were used as the OPD. 
The optical scale could not measure the actual OPD 
due to the thermal shrink of the scale and declination of 
the alignment between the optical scale and the optical path of each pixel. 
The correction of these effects 
is expected to be given by a set of constant scaling factors, which can be determined from well measured emission lines.
The observations of the Galactic Center and the Galactic Plane are 
suitable for determining 
these scaling factors because these sources are bright and have some strong 
emission lines in the wavenumber range of the FIS-FTS.

\begin{figure*}
\begin{center}
\FigureFile(80mm,80mm){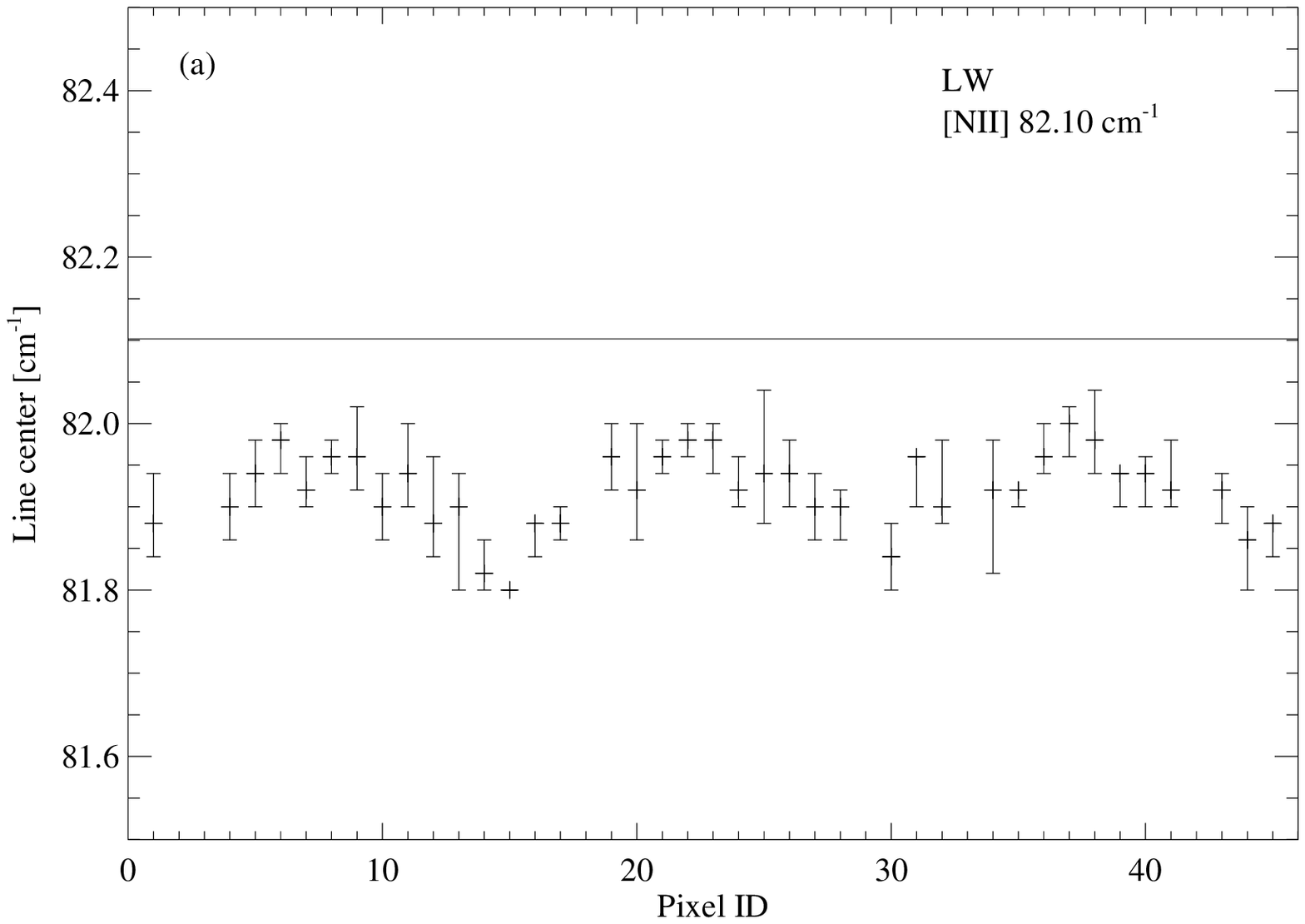}
\FigureFile(80mm,80mm){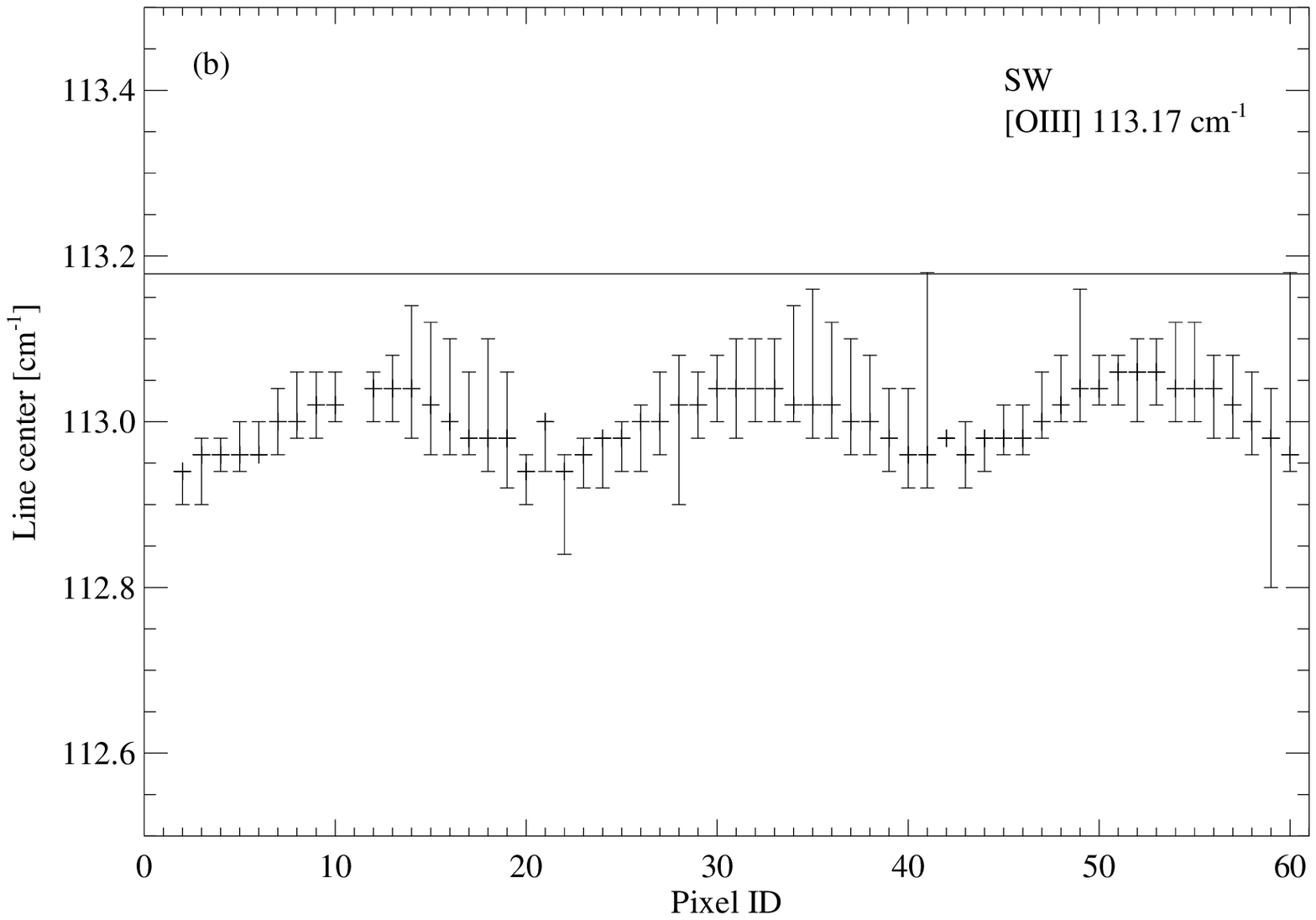}

\end{center}
\caption{Panel (a) and (b)  show the line centers of the best fit values of 
the [N\,{\scriptsize II}] line (82.10 cm$^{-1}$) and 
[O\,{\scriptsize III}] line (113.17 cm$^{-1}$), respectively, 

which are derived from the raw values of the optical scale.
}
\label{fig:line}
\end{figure*}

We fit the spectrum in the expected line profile and 
determine the line center in raw optical scale units.
The results are shown in figure \ref{fig:line}. 
Panel (a) and (b)  show the line centers of the best fit values of 
the [N\,{\footnotesize II}] line (82.10 cm$^{-1}$) and 
[O\,{\footnotesize III}] line (113.17 cm$^{-1}$), respectively.
Each data point indicates the median value and peak-to-peak variation of 
fitted line centers of all the observations (24 data points at maximum)
are indicated by error bar.

Rest frame wavenumbers of line centers are plotted by lines in each panel.
Due to the thermal shrink of the optical scale, all data points 
are located under the lines of expected wavenumbers.
We can see the common trend in figure \ref{fig:line} along the major axis 
of the detector arrays. 
This trend can be explained by the optical design of the FIS instrument as follows.  
The optical path of each pixel is not aligned 
with the optical axis; meanwhile, the optical scale is 
aligned with it.  
Therefore, the actual OPD of each pixel becomes 
longer than the measured length by the optical scale, which depends 
on the pixel position on the detector arrays.

Quantitatively, these shifts of line centers derived from the raw values 
of the optical scale are consistent with the expected values from the 
thermal shrink of the glass scale and the inclination of the optical path 
of each pixel.
The [C\,{\footnotesize II}] line (63.40 cm$^{-1}$) 
also shows a similar trend, but with larger scatter due to its 
wavenumber at the steep edge of the sensitivity change.

\section{Evaluation of the Calibration}

\subsection{Accuracy of the Calibration}

In this section, we summarize the absolute and relative accuracies 
for the continuum based on the results described in the previous sections.

As described in section 4.5, the absolute calibration accuracy 
of the FIS-FTS is $+20/-50\%$ for the total measured energy. 
In addition, the relative accuracy of the spectral response 
function is $\pm$10\% for entire spectral range of the SW.  
On the other hand, for LW it is $\pm5\%$ in 70--85 cm$^{-1}$  and several 
to 20\% in 60--70 cm$^{-1}$. Finally, we have to consider the 
accuracy of the flat correction, 10\%, described in 4.4. 

Considering all three uncertainties described above, the absolute 
accuracy is $+35/-55\%$ for SW, and $+35/-55\%$ in 70--85 cm$^{-1}$
and $+45/-60\%$ in 60--70 cm$^{-1}$ for LW. 
On the other hand, the relative accuracy 
among pixels of an array (the flat error) is $\pm$15\% for SW, 
and $\pm$10\% in 70--85 cm$^{-1}$ and $\pm$20\% in 60--70 cm$^{-1}$ for LW.

\subsection{Reproduction of some Spectra}

\begin{figure}
\begin{center}
\FigureFile(80mm,80mm){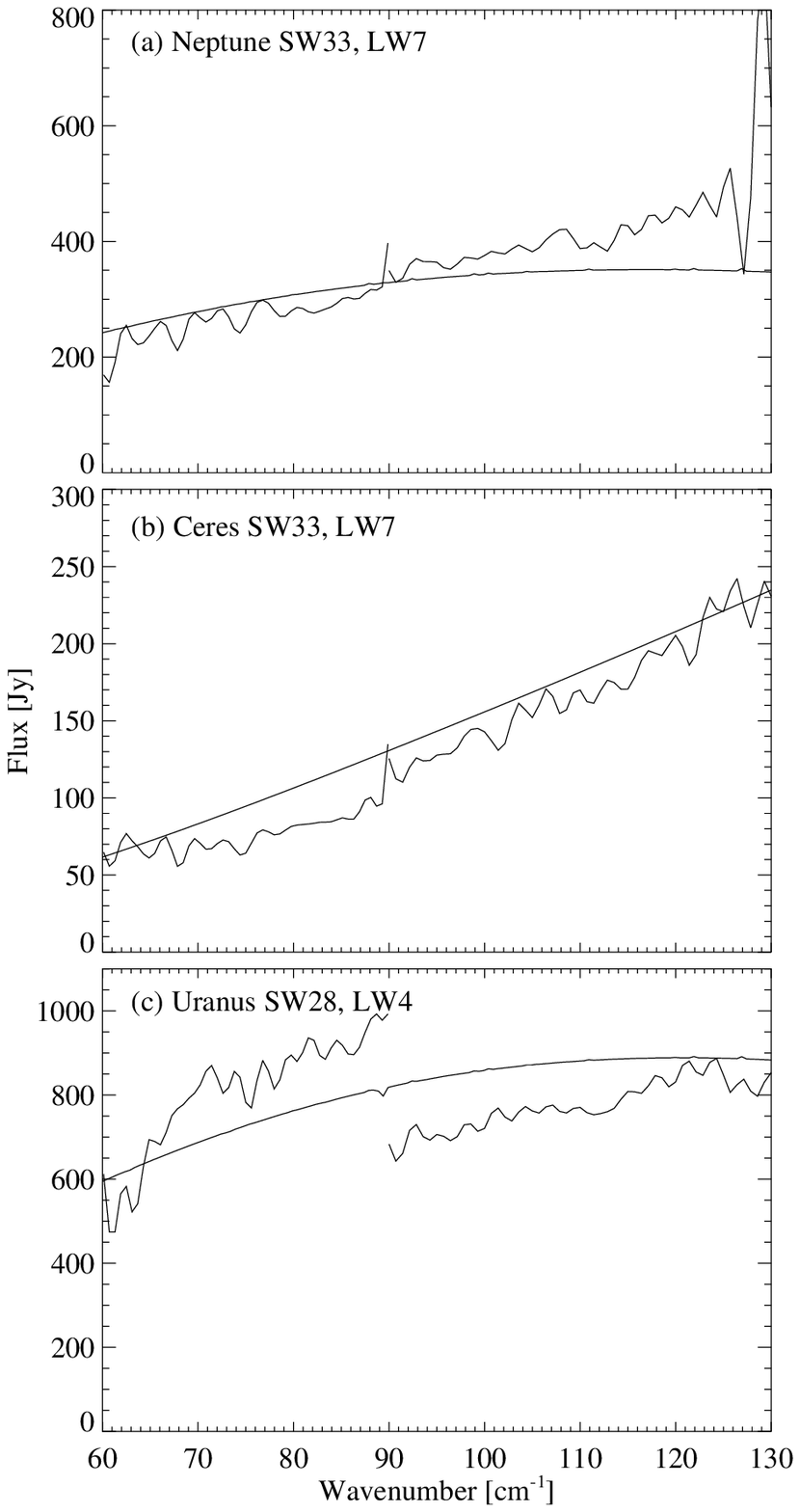}
\end{center}
\caption{Comparison between model spectra 
  (\cite{Muller98,Muller02,Moreno98}) and 
        calibrated spectra according to our procedure.  
	Three spectra are shown in panel (a) Neptune, 
	(b) Ceres, and (c) Uranus; Neptune and Ceres were observed with the 
	same position parameter (on-source pixel: SW33, LW7), and Uranus was 
	observed with a different position parameter (SW28, LW4).  
	These are reference 
	sources used for the spectral shape and/or absolute calibration.}
\label{fig:comp_solar}
\end{figure}

All the data used here are the same that were used 
in the calibration scheme. The evaluation of the spectra of planets
that are used 
to make the absolute calibration will confirm the self-consistency, and 
the evaluation using other data will test the reliability 
of the calibration method. 
The spectra of two planets, Uranus and Neptune, and a dwarf
planet, Ceres, have been derived according to the calibration
procedure described in this paper. These spectra are shown in
figure \ref{fig:comp_solar} together with empirical models. 
The derived spectra are in general agreement for
Neptune (a) and Ceres (b). However, a clear gap in the spectrum of
Uranus (c) between the SW and LW spectral range is observed.
 This gap, which is about 30\% of the flux level, 
is believed to arise from the uncertainty of the flat correction and 
position error of the source on the pixel. 
If the absolute flux is scaled to match the two spectral bands, 
the spectral shape is accurately reproduced with the model.

\section{Comparison with the ISO/LWS}

We compare the FIS-FTS spectra of the 
extended emission from the Galactic Center region, which is observed
in many pixels of the FIS-FTS and ISO/LWS.
The observational results of the ISO/LWS and FIS-FTS cannot be 
compared directly with each other, because of the differences of the 
beam sizes and the observing positions. 
Therefore, the spectral maps are constructed in RA-DEC coordinates 
from the FIS-FTS data first, 
and then compared with the ISO/LWS results. 
We investigate both continuum and line emissions.

\begin{figure}
\begin{center}
\FigureFile(80mm,80mm){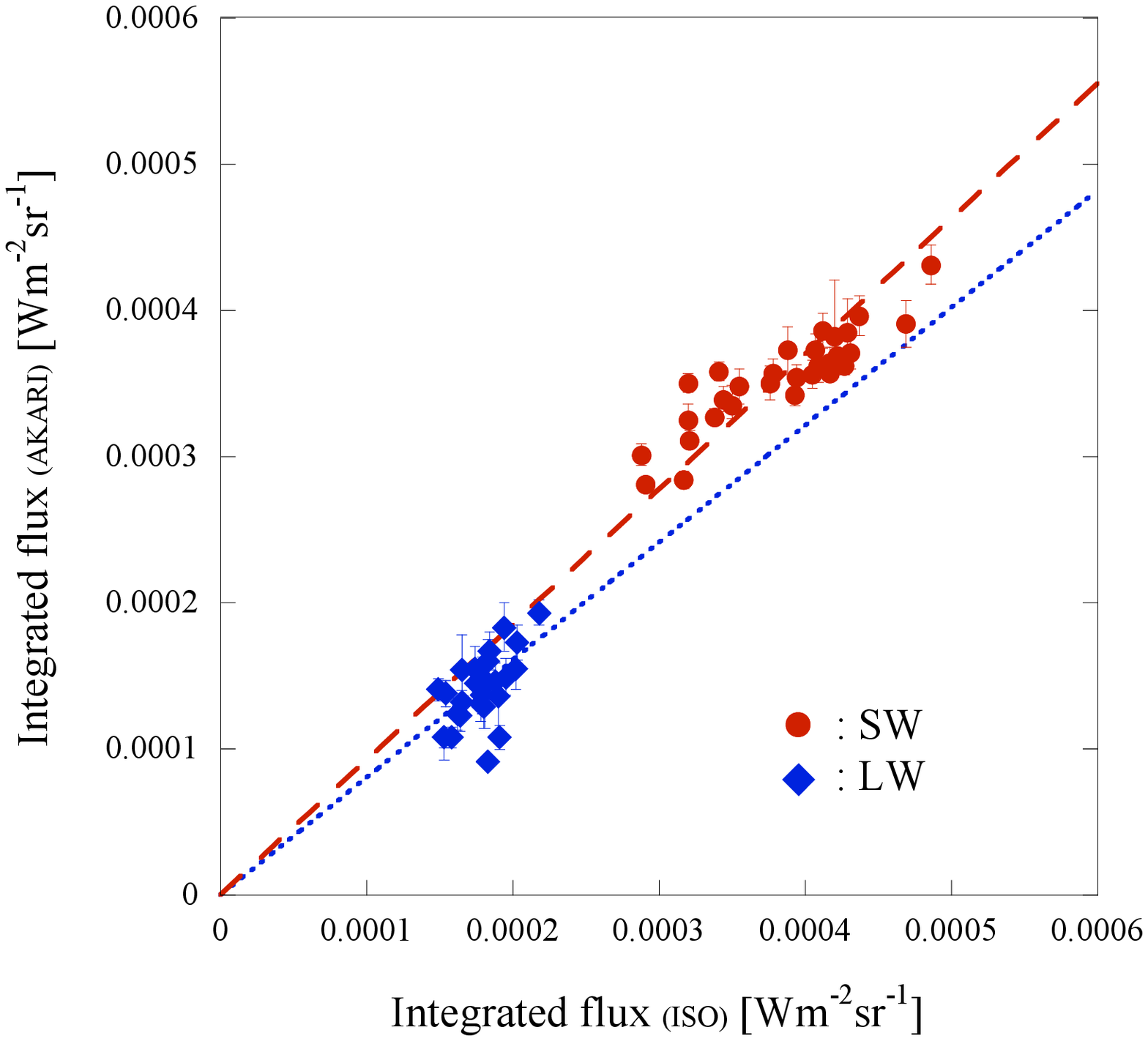}
\end{center}
\caption{Comparison of the ISO/LWS and the FIS-FTS integrated flux
of the continuum emission; the wavenumber range is the same
in each case. Circle and diamond symbols show data 
of SW and LW, respectively. The error bar of the FIS-FTS data is 
estimated from the discrepancy of forward and
backward scan data.
The dotted line and broken line represent the fitted slope for LW and
SW respectively (see text).}
\label{fig:akari_iso_cor_continuum}
\end{figure}

\begin{figure}
\begin{center}
\FigureFile(80mm,80mm){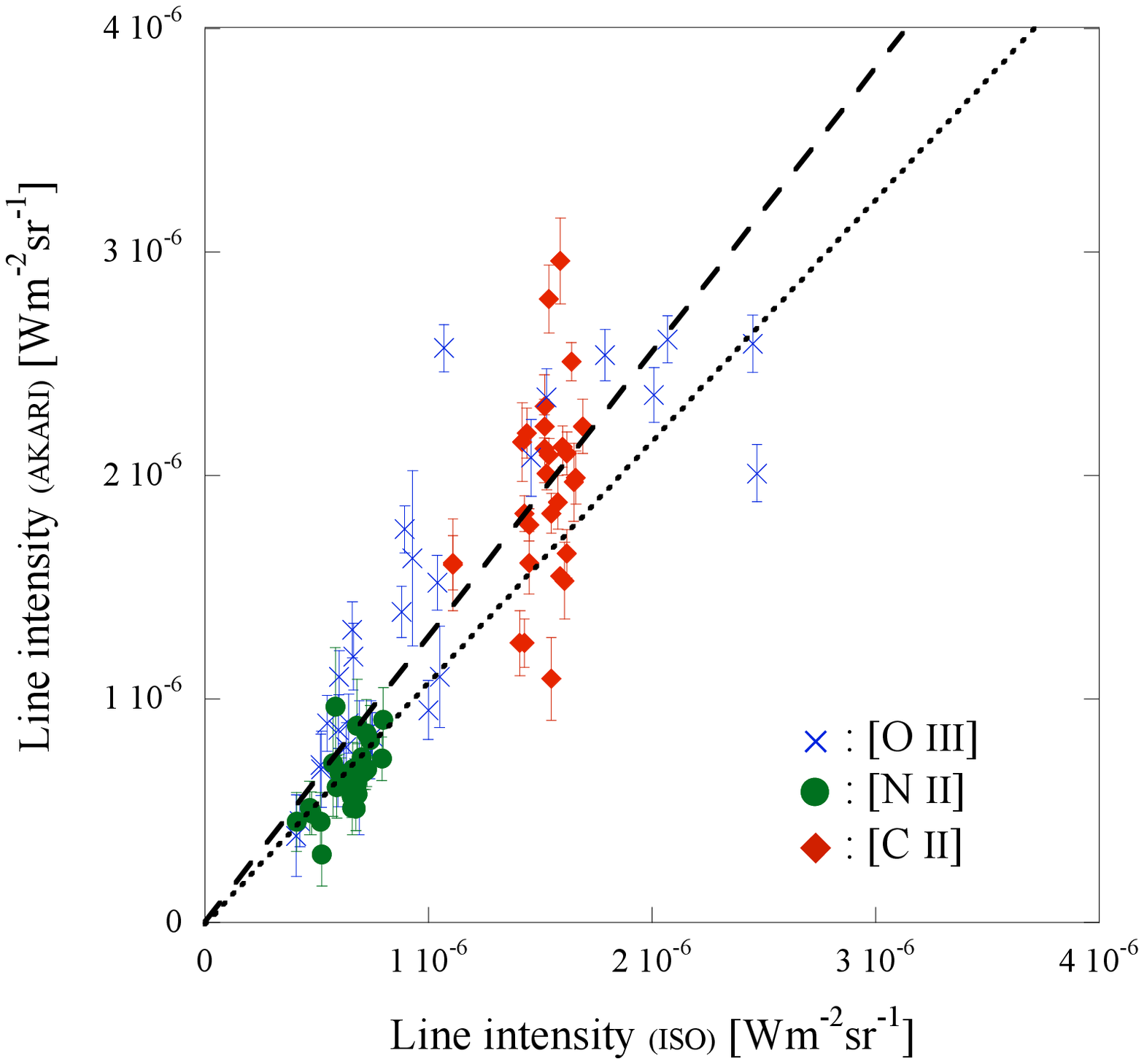}
\end{center}
\caption{Comparison of line intensities derived by 
the ISO/LWS and FIS-FTS. Diamond, circle, and cross symbols represent 
[C\,{\scriptsize II}] 63.40 cm$^{-1}$, [N\,{\scriptsize II}] 
82.10 cm$^{-1}$, 
and [O\,{\scriptsize III}] 113.17 cm$^{-1}$ lines respectively. 
The error bars of the FIS-FTS values indicate the fitting errors.
The dotted line represents the fitted slope for LW data
([C\,{\scriptsize II}] and [N\,{\scriptsize II}]).
On the other hand, the broken line represents SW data
([O\,{\scriptsize III}]) (see text).}
\label{fig:akari_iso_cor_line}
\end{figure}

To compare the continuum flux between the ISO/LWS and the FIS-FTS
measurements, integrated fluxes over the same spectral range (
65--85 cm$^{-1}$ for LW and 90--140 cm$^{-1}$ for SW),
excluding the line emission, are calculated for both
spectra. The point-to-point correlation of the integrated flux
between the ISO/LWS and the FIS-FTS spectra is shown in figure
\ref{fig:akari_iso_cor_continuum}.  

The ratio of the measured intensities of the FIS-FTS to those of the ISO/LWS 
is 0.93 on average with $\pm$0.07 of 1$\sigma$ error for SW and 0.79 on 
average with $\pm$0.11 of 1 $\sigma$ error for LW. 
The difference is within the 
absolute accuracy of the FIS-FTS, $+20/-50\%$, and that of the ISO/LWS (50\%) 
for an extended source (\cite{Gry03}). 
The scatter of the data 
points shown in figure \ref{fig:akari_iso_cor_continuum} 
is within the relative accuracy of the FIS-FTS, $\pm$15\% for SW 
and $\pm$10\% for LW. 

Figure~\ref{fig:akari_iso_cor_line} shows the comparison for the 
line intensities measured by the FIS-FTS and by the ISO/LWS. 
The measured lines are the fine structure lines of 
[O\,{\scriptsize III}] 113.17 cm$^{-1}$, [N\,{\footnotesize II}] 
82.10 cm$^{-1}$, and [C\,{\footnotesize II}] 63.40 cm$^{-1}$. 
The [O\,{\footnotesize III}] line appears in the SW 
spectral range, and other two lines are in LW. 
The average ratio of 
the [O\,{\footnotesize III}] intensity measured by the FIS-FTS to that of 
the ISO/LWS is 1.35 ($\pm$0.37 in 1$\sigma$ error), and 
[N\,{\footnotesize II}] and [C\,{\footnotesize II}] is 
1.15 ($\pm$0.27 in 1$\sigma$ error).
In contrast to the continuum, the line intensities measured by 
the FIS-FTS are systematically larger than those by the ISO/LWS, 
i.e., the line to continuum ratios are larger than the ISO/LWS.
This result may originate from the difference between Fourier 
transform spectrometers and grating spectrometers. In the case of Fourier 
transform spectrometers, the continuum intensity is dominantly 
determined from the modulation amplitude near the central burst, 
whereas the line intensities are determined from the periodic modulation 
of the over all interferogram. The Ge:Ga photoconductor may show
non-linear behavior caused by a transient response for a 
large signal near the central burst of the 
interferogram, resulting in relatively smaller signals around the central 
burst. Therefore, the continuum intensity measured by a Fourier 
transform spectrometer may produce smaller signals relative to the 
line intensity. On the other hand, in the case of grating spectrometers, 
the line components may diffract not only to 
the correct position but diffusely to all positions. 
Therefore, the line intensities measured by 
grating spectrometers tend to be smaller relative to the continuum 
intensities if the proper absolute calibrations are not made individually.

The scattering seen in figure~\ref{fig:akari_iso_cor_line} 
(for the lines) is much larger than that in 
figure~\ref{fig:akari_iso_cor_continuum} (for the continuum). 
It is larger than the relative accuracies described in the 
previous section, $\pm$15\% for SW, and $\pm$10\% in 70--85 cm$^{-1}$ 
and $\pm$20\% in 60--70 cm$^{-1}$ for LW.
For the [C\,{\footnotesize II}] line, the larger scatter is considered 
to occur because the line lies near the edge of the effective spectral 
range of the FIS-FTS.
The [O\,{\footnotesize III}] line tends to have large variation in 
smaller spatial scales than other lines, and thus, may be affected by the 
difference in the spatial resolution and the observed position 
of the sky between the FIS-FTS and ISO/LWS. 

In summary, the continuum intensities measured by the FIS-FTS are 
consistent with those measured by the ISO/LWS within their absolute accuracies, 
$+20/-50\%$ for the FIS-FTS and $\pm$50\%  for the ISO/LWS for extended sources
 (\cite{Gry03}).
The scatter seen in figure~\ref{fig:akari_iso_cor_continuum} can be 
explained by the flat error of the FIS-FTS. On the other hand, 
the line intensities measured by the FIS-FTS are systematically larger than 
those measured by the ISO/LWS compared to the continuum emission.
Based on the ISO/LWS, the line to continuum ratio of the FIS-FTS may be overestimated 
by 45\% for both of SW and LW.
The scatter seen in 
figure~\ref{fig:akari_iso_cor_line} is larger than that expected 
from the calibration error, and can be explained by small differences in 
the spatial resolution as well as in the observing position.

\section{Conclusion}

FIS-FTS operated successfully after the launch of the AKARI
satellite until the supply of liquid helium expired. During its 1.5
year lifetime, the FIS-FTS acquired about 600 pointed observations. 
The properties of the imaging FTS
equipped with photoconductive detector arrays have been presented. 
The calibration method discussed is
based on observations of bright astronomical sources such as
planets, dwarf planets, and extended sources in our Galaxy, as well
as instrumental sources such as the internal blackbody source and
the aperture lid.  The current version of the calibration method
ignores the transient response of detectors, though its effect can
be seen as a distortion of the interferograms. 
Under this condition the relative uncertainty of the  calibration of the 
continuum 
is estimated to be $\pm$15\% for SW, and $\pm$10\% in 70--85 cm$^{-1}$ 
and $\pm$20\% in 60--70 cm$^{-1}$ for LW, 
and the absolute uncertainty is estimated to be $+35/-55\%$ for SW, 
 and $+35/-55\%$ in 70--85 cm$^{-1}$ 
and $+40/-60\%$ in 60--70 cm$^{-1}$ for LW.
These values are confirmed by comparison with theoretical models and 
previous observations by the ISO/LWS.

It should be noted that the calibration of the FIS-FTS 
is limited to bright sources. It is known
that the transient response of detectors depends on both the
background and the source fluxes. 
It has not been established whether
the calibration presented in this paper can be extended to faint
sources.  
A more
detailed calibration study will tackle the challenging
problem of the detector transient response and correct the
interferograms in the time domain before the Fourier transform. It is
expected that the calibration accuracy will be improved through
the correction of the transient response of the detector (\cite{Kaneda09}) and 
the use of the
larger number of lower brightness sources observed with the FIS-FTS.

\bigskip

AKARI is a JAXA project with the participation of ESA.
We thank all the members of the AKARI project for their continuous help
and support. The FIS was developed in collaboration with Nagoya
University, ISAS, the University of Tokyo, the National Institute of Information
and Communications Technology (NICT), the National Astronomical
Observatory of Japan (NAOJ), and other research institutes. 
We thank ESA for financial support. We also
thank Prof. R. Moreno, and Prof. T. G. Mueller for calculating the 
model of the
solar system objects which were used for the absolute calibration.

%\begin{figure}
%  \begin{center}
%    \FigureFile(80mm,80mm){fig1.eps}
%    %%% \FigureFile(width,height){filename}
%  \end{center}
%  \caption{This is the first figure.}\label{fig:sample}
%\end{figure}

%%%%%%%%%%%%%%%%%%%%%%%%%%%%%%%%%%%%%%%

%\begin{table}
%  \caption{This is the first tabular.}\label{tab:first}
%  \begin{center}
%    \begin{tabular}{llll}
%      \hline
%      a & b & c & d \\
%      e & f & g & h \\
%      ....\\
%      \hline
%    \end{tabular}
%  \end{center}
%\end{table}

%\begin{longtable}{lll}
%  \caption{Sample of ``longtable"}\label{tab:LTsample}
%  \hline              
%  name & value1 & value2 \\ 
%\endfirsthead
%  \hline
%  name & value & value2  \\
%\endhead
%  \hline
%\endfoot
%  \hline
%\endlastfoot
%  \hline
%  aaaaa & bbbbb & ccccc \\
%  ...... & ..... & ..... \\
%  ...... & ..... & ..... \\
%  ...... & ..... & ..... \\
%  xxxxx & yyyyy & zzzzz \\
%\end{longtable}

%\appendix
%\section{Method of .....}

%\section{Approximation of ...}

%\section*{Complete data}

%%%
% See the manual for the detail.
%%%

\end{document}